\documentclass[3p]{elsarticle}
\usepackage{graphicx}
\usepackage{amssymb,amsmath,color}
\pdfoutput=1

\newcommand{\D}[2]{\frac{\partial #2}{\partial #1}}

\newcommand{\deriv}[2]{\frac{{\rm d} #2}{{\rm d} #1}}

\newcommand\bb[1]{\mbox{\boldmath{$#1$}}}
\newcommand\grad{\bb{\nabla}}
\newcommand\bcdot{\bb{\cdot}}
\newcommand\btimes{\bb{\times}}
\newcommand{\mc}[1]{\mathcal{#1}}

\newcommand{\msb}[1]{\bb{\mathsf{#1}}}

\newcommand{\imag}{{\rm i}}
\newcommand{\ex}{\hat{\bb{e}}_x}
\newcommand{\ey}{\hat{\bb{e}}_y}
\newcommand{\ez}{\hat{\bb{e}}_z}
\newcommand{\eb}{\hat{\bb{b}}}

\begin{document}

\begin{frontmatter}

\title{{\it Pegasus}: A New Hybrid-Kinetic Particle-in-Cell \\ Code for Astrophysical Plasma Dynamics}

\author[pu,mppc]{Matthew W. Kunz\corref{cor}\fnref{einstein}}
\author[pu,mppc]{James M. Stone}
\author[itc]{and Xue-Ning Bai\fnref{hubble}}

\address[pu]{Department of Astrophysical Sciences, Princeton University, Princeton, NJ 08544, United States}
\address[mppc]{Max-Planck/Princeton Center for Plasma Physics}
\address[itc]{Institute for Theory and Computation, Harvard-Smithsonian Center for Astrophysics, Cambridge, MA 02138, United States}
\cortext[cor]{Corresponding author: mkunz@princeton.edu; +1 609 258 1903}
\fntext[einstein]{NASA Einstein Postdoctoral Fellow.}
\fntext[hubble]{NASA Hubble Postdoctoral Fellow.}

%
% Abstract
%
\begin{abstract}
We describe {\em Pegasus}, a new hybrid-kinetic particle-in-cell code tailored for the study of astrophysical plasma dynamics. The code incorporates an energy-conserving particle integrator into a stable, second-order--accurate, three-stage predictor-predictor-corrector integration algorithm. The constrained transport method is used to enforce the divergence-free constraint on the magnetic field. A $\delta f$ scheme is included to facilitate a reduced-noise study of systems in which only small departures from an initial distribution function are anticipated. The effects of rotation and shear are implemented through the shearing-sheet formalism with orbital advection. These algorithms are embedded within an architecture similar to that used in the popular astrophysical magnetohydrodynamics code {\it Athena}, one that is modular, well-documented, easy to use, and efficiently parallelized for use on thousands of processors. We present a series of tests in one, two, and three spatial dimensions that demonstrate the fidelity and versatility of the code.
\end{abstract}


\end{frontmatter}

%
% Introduction
%
\section{Introduction}\label{sec:intro}

Many astrophysical systems, such as hot accretion flows, the intracluster medium of galaxy clusters, the solar wind, and some phases of the interstellar medium, are weakly collisional and magnetized. As a result, their mean, global properties are vastly separated in both space and time from the detailed kinetic microphysics that governs the transport and dissipation of momentum, energy, and magnetic fields. As these systems are more often than not highly nonlinear, one is faced with the daunting task of developing a rigorous numerical approach that can simultaneously grapple with these nonlinearities while respecting the scale hierarchy.

Currently, the two most common strategies for overcoming this hurdle in the astrophysical community are the use of fully kinetic particle-in-cell (PIC) codes and the construction of sub-grid models for use in magnetohydrodynamic (MHD) codes. The advantage of the former is that it rigorously solves for everything: Debye-scale plasma oscillations, electron and ion Larmor-scale physics, the propagation of electromagnetic waves, collective macroscale dynamics, etc. This is also a disadvantage. The systems mentioned above are non-relativistic, quasi-neutral, hydrogenic plasmas. Not only is one uninterested in describing the propagation of light or the emergence of quasi-neutrality in these systems, but also the latter property (hydrogenic) means that the electron and proton scales are well-separated. Furthermore, the expense of a full electromagnetic PIC code more often than not demands a reduced speed of light and/or a depressed ion-to-electron mass ratio, corrupting the scale hierarchy and entangling physics that ought to be spatially and temporally well-separated. On the other hand, one can avoid simulating this scale hierarchy by simply describing the plasma with a set of magnetohydrodynamic (MHD) or kinetic-MHD equations, coupled with a sub-grid closure that mimics the effects of microscale phenomena on the macroscale dynamics. This has been a profitable approach, used in numerical studies of the collisionless MRI \cite{shqs06} and of thermal convection in weakly collisional plasmas \cite{kbrs12}. But one is constantly reminded that the assumptions built into such closures have not been rigorously justified.

One compromise is the hybrid-kinetic model \cite{bcch78,hn78}. Non-relativistic electrons are taken to be a massless, neutralizing, MHD-like fluid, while the ions are treated kinetically. This eliminates Debye-scale physics, plasma oscillations, and the speed of light, while guaranteeing scale separation and capturing crucial microscale phenomena. Consequently, the resulting system of equations are much cheaper to solve numerically. This is the approach we have adopted in building {\it Pegasus}, which solves the hybrid-kinetic equations using a PIC approach.

Hybrid-kinetic PIC codes have a long history \cite{wo93,lipatov02,wyokq03}, one that is often tied to solving problems in space physics and laboratory plasmas. They have also been used to study particle acceleration and magnetic-field amplification in collisionless astrophysical shocks. Our goal in building {\it Pegasus} is to extend this tradition even further into the domain of astrophysical plasmas. Accordingly, gravity is allowed for in the code, as are rotation and shear through the shearing-sheet formalism and a technique known as orbital advection. Several different boundary conditions are included, including inflow boundaries with particle injection. A $\delta f$ scheme is made available to facilitate a reduced-noise study of systems in which only small departures from an initial distribution function are expected. All of these features are embedded within an architecture similar to that used in the popular astrophysical MHD code {\em Athena} \cite{sgths08}, one that is modular, well-documented, easy to use, and efficiently parallelized for use on thousands of processors.

There are also several features that distinguish the integration scheme designed for {\it Pegasus} from those used in many other hybrid-kinetic PIC codes. In {\it Pegasus}, the magnetic field is updated using constrained transport, which preserves the solenoidal character of the magnetic field to machine precision. Our three-stage time integration scheme, based on the Crank-Nicholson method, is second-order--accurate, can stably propagate Alfv\'{e}n and whistler waves at all scales, is total-variation-diminishing, and exhibits excellent conservation properties. It takes advantage of modern advances in computing power and memory to employ prediction-stage particle updates (rather than forward-in-time linear extrapolation or a moment method) to help determine time-advanced values of the electric and magnetic fields. While describing these features, we point out that whistler waves are numerically unstable in unsplit integration schemes with only two steps, including second-order Runge-Kutta and Lax-Wendroff. We also caution that, for some iterative predictor-corrector schemes commonly used in hybrid-kinetic codes, numerical stability is not always guaranteed as the number of iterations is increased.

The outline of the paper is as follows. In Section \ref{sec:equations}, we formulate the problem of solving the Vlasov-Landau kinetic equation for fluid electrons and kinetic ions. The shearing-sheet formalism is discussed, and the dimensionless free parameters in the problem are given. In Section \ref{sec:code}, we present each of the ingredients of our code, closing with a description of the integration scheme, its implementation in {\it Pegasus}, and its performance on modern parallel computers. Having described the code in detail, we then present in Section \ref{sec:tests} the results of some representative tests we have performed to demonstrate the accuracy, stability, and versatility of our code. Finally, in Section \ref{sec:summary} we close with a summary of the code and a brief description of how it is currently being used to solve several modern problems in astrophysics and space science.

%
% Equations
%
\section{Hybrid-Kinetic Equations}\label{sec:equations}

\subsection{Basic equations}

The distribution function $f_s ( t, \bb{r} , \bb{v} )$ satisfies the Vlasov-Landau kinetic equation
\begin{equation}\label{eqn:vlasov}
\D{t}{f_s} + \bb{v} \bcdot \grad f_s + \left[ \frac{q_s}{m_s} \left( \bb{E} + \frac{\bb{v}}{c} \btimes \bb{B} \right) + \bb{g} \right] \bcdot \frac{\partial f_s}{\partial\bb{v}} = C [ f_s ] ,
\end{equation}
where $s = {\rm i},{\rm e}$ is the particle species, $\bb{r}$ is its position, $\bb{v}$ is its velocity, and $q_s$ and $m_s$ are its charge and mass ($q_{\rm e} = -e$, $q_{\rm i} = Ze$). The term on the right-hand side is the collision operator, and $\bb{g}$ denotes the gravitational acceleration (or, more generally, any velocity-independent force per unit mass). The electric field $\bb{E}$ and magnetic field $\bb{B}$ are determined from Maxwell's equations: quasi-neutrality
\begin{equation}\label{eqn:quasineutrality}
\sum_s q_s n_s \equiv \sum_s q_s \int {\rm d}^3 \bb{v} \, f_s = 0 ,
\end{equation}
where $n_s$ is particle number density; Amp\'{e}re's law
\begin{equation}\label{eqn:current}
\bb{j} = \sum_s q_s n_s \bb{u}_s \equiv \sum_s q_s \int {\rm d}^3 \bb{v} \, \bb{v} f_s = \frac{c}{4\pi} \grad \btimes \bb{B} ,
\end{equation}
where $\bb{j}$ is the current density and $\bb{u}_s$ is the mean velocity of species $s$; Faraday's law
\begin{equation}\label{eqn:faraday}
\D{t}{\bb{B}} = -c \grad \btimes \bb{E} ;
\end{equation}
and $\grad \bcdot \bb{B} = 0$. 

In writing equations (\ref{eqn:quasineutrality}) and (\ref{eqn:current}), we are restricting our attention to scales larger than the Debye length and velocities much less than the speed of light. This restriction allows us to perform Galilean transforms on these equations. For example, it will often be convenient to calculate the distribution function in terms of peculiar velocities $\bb{v}' = \bb{v} - \bb{u}_0(t,\bb{r})$ relative to a specified velocity field $\bb{u}_0(t,\bb{r})$. Transforming the variables $(t,\bb{r},\bb{v}) \rightarrow (t, \bb{r}, \bb{v}')$, equation (\ref{eqn:vlasov}) takes the form
\begin{equation}\label{eqn:vlasov-peculiar}
\D{t}{f_s} + \bb{u}_0 \bcdot \grad f_s + \bb{v}' \bcdot \grad f_s + \left[ \frac{q_s}{m_s} \left( \bb{E}' + \frac{\bb{v}'}{c} \btimes \bb{B} \right) + \bb{g}
- \D{t}{\bb{u}_0} - \bb{u}_0 \bcdot \grad \bb{u}_0 - \bb{v}' \bcdot \grad \bb{u}_0 \right] \bcdot \frac{\partial f_s}{\partial\bb{v}'} = C [ f_s ] ,
\end{equation}
where $\bb{E}' = \bb{E} + \bb{u}_0 \btimes \bb{B} / c$ is the electric field in the co-moving frame. Likewise, equation (\ref{eqn:faraday}) becomes
\begin{equation}\label{eqn:faraday-peculiar}
\D{t}{\bb{B}} = -c \grad \btimes \bb{E}' + \grad \btimes \left( \bb{u}_0 \btimes \bb{B} \right) .
\end{equation}
This separation will prove useful for problems in which the particles' velocities have an analytically separable piece---one that we would rather not entrust a finite number of particles with representing. One notable example is the shearing sheet, which we now describe.

\subsection{Shearing-sheet formalism}

When studying rotating and shearing plasmas, we adopt a framework commonly referred to as the shearing sheet \cite{glb65,hgb95}. We consider equations (\ref{eqn:vlasov})--(\ref{eqn:faraday}) in a local Cartesian reference frame that co-orbits with a fixed position $r_0$ in the unperturbed flow at an angular velocity $\bb{\varpi} = \varpi_0 \ez$. The $x$, $y$, and $z$ directions correspond, respectively, to the local radial, azimuthal, and vertical directions. Differential rotation is accounted for by including the Coriolis and centrifugal forces, and by imposing a background linear shear, $\bb{u}_0(x) = -\sigma_0 x\,\ey$, where $\sigma_0 \equiv -r_0 ( {\rm d} \varpi / {\rm d} r )_{r=r_0}$; Keplerian rotation yields $\sigma_0 = (3/2) \varpi_0$.\footnote{In much of the accretion-disk literature, the shear frequency $\sigma_0$ is written as $q \varpi_0$. Here we retain a more general form for the shear frequency, so that the case of non-zero shear may be considered in non-rotating systems.} Then, Equation (\ref{eqn:vlasov-peculiar}) becomes
\begin{equation}\label{eqn:ssvlasov}
\left( \D{t}{} - \sigma_0 x \D{y}{} \right) f_s + \bb{v}' \bcdot \grad f_s + \left[ \frac{q_s}{m_s} \left( \bb{E}' + \frac{\bb{v}'}{c} \btimes \bb{B} \right) + \bb{g} - \varpi^2_0 \, \ez \btimes ( \ez \btimes \bb{r} ) - 2 \varpi_0 \ez \btimes \bb{v}' + \sigma_0 v_x' \ey  \right] \bcdot \frac{\partial f_s}{\partial\bb{v}'} = C [ f_s ] .
\end{equation}
In the standard shearing-sheet formalism, the (radial) gravitational acceleration is exactly offset by the centrifugal acceleration, so that $\bb{g} = \varpi^2_0 \, \ez \btimes ( \ez \btimes \bb{r} )$. In the shearing frame, equation (\ref{eqn:faraday-peculiar}) becomes
\begin{equation}\label{eqn:ssinduction}
\left( \D{t}{} - \sigma_0 x \D{y}{} \right) \bb{B} = - c \grad \btimes \bb{E}' - \sigma_0 B_x \ey .
\end{equation}
The second terms on the left-hand sides of equations (\ref{eqn:ssvlasov}) and (\ref{eqn:ssinduction}) represent the advection of particles and magnetic flux, respectively, by the background shear flow. The final terms in these equations represent the stretching of the velocity and magnetic field lines in the $y$ direction by the background shear.

For full generality, we retain $\varpi_0$ and $\sigma_0$ in everything that follows; applications to plasmas not being studied within the shearing-sheet formalism can be readily treated by setting $\varpi_0 = \sigma_0 = 0$ in the appropriate equations. Henceforth, we drop the primes on $\bb{v}'$ and $\bb{E}'$. 

\subsection{Electron and ion equations}

For the electron species ($s = {\rm e}$), we solve equation (\ref{eqn:ssvlasov}) by expanding the electron distribution function in the square root of the electron-ion mass ratio $(m_{\rm e} / m_{\rm i} )^{1/2} \approx 0.02$, a natural small parameter for the plasma (e.g.~see Appendix A1 of ref.~\cite{rsrc11}). The outcome of the mass-ratio expansion is that the electrons are Maxwellian, gyrotropic, and isothermal ($T_{\rm e} = {\rm const}$), and that the electric field can be written in terms of $\bb{u}_{\rm e}$, $\bb{B}$, and $n_{\rm e}$ via a generalized Ohm's law:
\begin{equation}
\bb{E} + \frac{\bb{u}_{\rm e}}{c} \btimes \bb{B} - \frac{\eta}{c} \grad \btimes \bb{B} = - \frac{\grad p_{\rm e}}{e n_{\rm e}} = -\frac{T_{\rm e} \grad n_{\rm e}}{e n_{\rm e}} ,
\end{equation}
where we have adopted a simple model for the electron collision operator by introducing the magnetic diffusivity $\eta$. This can be recast in terms of the mean velocity of the ions $\bb{u}_{\rm i}$, the number density of ions $n_{\rm i}$, and $\bb{B}$ using equations (\ref{eqn:quasineutrality}) and (\ref{eqn:current}):
\begin{equation}\label{eqn:efield}
\bb{E} + \frac{\bb{u}_{\rm i}}{c} \btimes \bb{B} - \frac{\eta}{c} \grad \btimes \bb{B} = - \frac{T_{\rm e} \grad n_{\rm i}}{ e n_{\rm i}} + \frac{ ( \grad \btimes \bb{B} ) \btimes \bb{B} }{ 4 \pi Z e n_{\rm i} } .
\end{equation}
The first term on the right-hand side of this equation represents the thermoelectric effect.\footnote{Polytropic electrons can be described by letting $T_{\rm e} \rightarrow \Gamma_{\rm e} T_{\rm e}$ in equation (\ref{eqn:efield}) with $T_{\rm e} \propto n_{\rm i}^{\Gamma_{\rm e} - 1}$.}  The second term is the Hall electric field, generated by the differential motion between ions and electrons. Note that, for barotropic electrons, the thermoelectric field does not break magnetic-field lines.

For the ion species ($s = {\rm i}$), equation (\ref{eqn:ssvlasov}) is solved using the method of characteristics. For scales at which ion--ion collisions are negligible, the ion distribution function $f_{\rm i} (t,\bb{r}, \bb{v})$ is constant on trajectories defined by
\begin{subequations}\label{eqn:characteristics}
\begin{equation}
\deriv{t}{\bb{x}} = \bb{v} - \sigma_0 x \ey \qquad {\rm and} \qquad \deriv{t}{\bb{v}} = \frac{Ze}{m_{\rm i}} \left( \widetilde{\bb{E}} + \frac{\bb{v}}{c} \btimes \bb{B} \right) - 2 \varpi_0 \ez \btimes \bb{v} + \sigma_0 v_x \ey ,
\end{equation}
where
\begin{equation}
\widetilde{\bb{E}} \equiv \bb{E} - \frac{\eta}{c} \grad \btimes \bb{B} + \frac{m_{\rm i}}{Ze} \Bigl[ \bb{g} - \varpi^2_0 \, \ez \btimes ( \ez \btimes \bb{r} ) \Bigr]
\end{equation}
\end{subequations}
is the effective electric field felt by the particles.\footnote{While there are a wealth of ideas on how to incorporate Coulomb collisions into PIC codes \cite{ta77,dc94,mc94,jones96,mlj97,larson03,cp07,sherlock08,lwda09,ht10}, we defer to a future publication a discussion of their inclusion in {\em Pegasus}. For now we simply remark that, in $\delta f$ simulations (\S\ref{sec:deposit}) without explicit collisions, there is a ``growing weights problem" \cite{kh94,krommes99,cp07} for simulations in which a free-energy source is maintained (as in the shearing sheet).}

Equations (\ref{eqn:quasineutrality}), (\ref{eqn:current}), (\ref{eqn:ssinduction}), (\ref{eqn:efield}), and (\ref{eqn:characteristics}) constitute the hybrid kinetics.

\subsection{Natural units and dimensionless free parameters}

The equations are put in dimensionless form by choosing units natural to the system. The units of velocity $[v]$, time $[t]$, and magnetic-field strength $[B]$ are, respectively, the initial Alfv\'{e}n speed in the ions, $v_{\rm A,0} \equiv B_0 / ( 4 \pi m_{\rm i} n_{\rm i,0} )^{1/2}$; the inverse of the initial Larmor frequency of the ions, $\Omega^{-1}_{\rm i,0} \equiv m_{\rm i} c / Z e B_0$; and the initial magnetic-field strength, $B_0$. The implied unit of length $[L]$ is the initial ion skin depth $d_{\rm i,0}$.

In these units, the governing equations have four dimensionless free parameters: the initial plasma beta of the ions, $\beta_0 \equiv 8 \pi n_{\rm i,0} T_{\rm i,0} / B^2_0 = v^2_{\rm th,i,0} / v^2_{\rm A,0}$; the ratio of the initial ion and electron temperatures, $\tau \equiv T_{\rm i,0} /  Z T_{\rm e,0}$; the ratio of the rotation and initial ion Larmor frequencies, $\varpi \equiv \varpi_0 / \Omega_{\rm i,0}$; and the ratio of the shear and initial ion Larmor frequencies, $\sigma \equiv \sigma_0 / \Omega_{\rm i,0}$. If an effective gravity and magnetic diffusivity are included, they are measured in units of $v_{\rm A,0} \Omega_{\rm i,0}$ and $v_{\rm A,0} d_{\rm i,0}$, respectively.

%
% Code
%
\section{Numerical Approach}\label{sec:code}

We solve the hybrid-kinetic equations using a particle-in-cell approach. The continuous set of characteristics defined by equation (\ref{eqn:characteristics}) are replaced by a discrete subset: $N_{\rm p}$ finite-sized particles with positions $\bb{x}_p$ and velocities $\bb{v}_p$, where $p=1\dots N_{\rm p}$ is the index of the $p$th particle, are integrated forward in time according to equation (\ref{eqn:characteristics}). Along these characteristics, the particles are deposited onto a uniform Cartesian grid to calculate the zeroth (eq.~\ref{eqn:quasineutrality}) and first (eq.~\ref{eqn:current}) moments of the ion distribution function. With the ion density and momentum discretized, the electric field can be calculated on the grid using equation (\ref{eqn:efield}). The magnetic field is then advanced using Faraday's law (eq.~\ref{eqn:ssinduction}).

Rendering this process numerically accurate and stable requires the adoption of several specialized algorithms. First, one must construct a (preferably symplectic) method for updating (``pushing") the particles. There must also be a means of accurately depositing these particles onto the computational grid to calculate moments of the ion distribution function, as well as an algorithm for reducing the influence of finite-number particle noise on this calculation. The magnetic field must be updated in such a way as to guarantee the stable propagation of waves while preserving the solenoidality constraint $\grad \bcdot \bb{B} = 0$. When working in the shearing sheet, further measures must be taken to enforce shearing periodicity and to guarantee the stable and accurate advection of particles and fields by the background shear flow. In this Section we detail the algorithms constructed to meet each of these requirements and provide a discussion of how these ingredients are combined in {\it Pegasus} to perform one time-step.

In what follows, the number density $n$, the mean velocity $\bb{u}$, and the pressure $\msb{P}$ all refer to the ion species, as do the particle positions $\bb{x}$ and velocities $\bb{v}$. To avoid confusion between what will be the time-step label $n$ and the ion number density $n$, the time-step label is always written as a superscript in parentheses (e.g.~$t^{(n)}$).

\subsection{Modified Boris push in the shearing sheet}\label{sec:boris}

We begin the description of our numerical integration algorithm with the ion-particle push. We integrate equation (\ref{eqn:characteristics}) from time $t^{(n)}$ to time $t^{(n+1)}$ using a semi-implicit scheme based on the Boris algorithm \cite{boris70}. Given the electric and magnetic fields at time $t^{(n+1/2)}$ (\S\ref{sec:algorithm}), the basic algorithm is:
\begin{subequations}\label{eqn:push}
\begin{align}
\bb{x}^\ast& = \bb{x}^{(n)} + \frac{h}{2} \, \bb{v}^{(n)}  \\*
\bb{v}^-& = \bb{v}^{(n)} +  \frac{h}{2} \, \widetilde{\bb{E}}^{(n+1/2)} \\*
\bb{v}^+& = \bb{v}^-  + h\, \bb{a} \left( \frac{ \bb{v}^- + \bb{v}^+ }{2} , \bb{B}^{(n+1/2)}  \right) \label{eqn:vupdate} \\*
\bb{v}^{(n+1)} & = \bb{v}^+ +  \frac{h}{2} \, \widetilde{\bb{E}}^{(n+1/2)} \\*
\bb{x}^{(n+1)} & = \bb{x}^\ast +  \frac{h}{2} \, \bb{v}^{(n+1)} ,
\end{align}
\end{subequations}
where $h \equiv t^{(n+1)} - t^{(n)}$ is the size of the time-step and
\begin{equation}
\bb{a}( \bb{v} ,  \bb{B} ) \equiv \bb{v} \btimes \bb{B} - 2 \varpi \ez \btimes \bb{v} + \sigma v_x \ey
\end{equation}
is the velocity-dependent part of the acceleration. The electric, magnetic, and gravitational fields are evaluated at the predicted particle position $\bb{x}^\ast$ using second-order--accurate interpolation (\S\ref{sec:deposit}). This scheme is semi-implicit because the velocity update (\ref{eqn:vupdate}) depends on both $\bb{v}^-$ and $\bb{v}^+$ through Crank-Nicholson differencing. Converting to explicit form, equation (\ref{eqn:vupdate}) becomes
\begin{equation}\label{eqn:modboris}
\bb{v}^+ = \bb{v}^- + h \, \msb{M}^{-1} \bb{a} \left( \bb{v}^- , \bb{B}^{(n+1/2)} \right) , \quad {\rm where} \quad \msb{M} \equiv \msb{I} - \frac{h}{2} \D{\bb{v}}{\bb{a}} ,
\tag{\ref{eqn:vupdate}$'$}
\end{equation}
$\msb{I}$ is the unit dyadic, and the Jacobian
\begin{equation}
\D{\bb{v}}{\bb{a}} = 
\begin{pmatrix}
0 & B_z + 2 \varpi & - B_y \\
- B_z + \sigma - 2 \varpi & 0 & B_x \\
B_y & - B_x & 0
\end{pmatrix} .
\end{equation}
When $\sigma = \varpi = 0$, equation (\ref{eqn:modboris}) reduces to the standard Boris rotation \cite{boris70}; for uncharged particles in the shearing sheet, it reduces to the semi-implicit integrator introduced by Bai and Stone \cite{bs10}. After equations (\ref{eqn:push}) are performed, the $y$-position of the particle is shifted by an amount $-h \sigma \bar{x}$, where $\bar{x} \equiv ( \bb{x}^{(n)} + \bb{x}^{(n+1)} ) / 2$. In Section \ref{sec:orbit}, we show that this algorithm conserves energy to machine precision.

\subsection{Particle deposits in full-$f$ and $\delta f$ methods}\label{sec:deposit}

In order to calculate the electric field (eq.~\ref{eqn:efield}), the zeroth and first moments of the ion distribution function must be evaluated. This is done by depositing the particles onto the numerical grid. To do so, each particle is described by a shape function $S ( \bb{r} - \bb{x}_p )$, which describes the contribution of a particle at position $\bb{x}_p$ to a cell-centered quantity at the location $\bb{r}$ (see Fig.~\ref{fig:grid}). We employ the second-order--accurate triangular-shaped cloud (TSC; \cite{bl91}). The same shape function is used to interpolate the electric, magnetic, and gravitational fields to the locations of the particles for the modified Boris push (\S\ref{sec:boris}).\footnote{Special care is taken to insure that, at each particle location, the interpolated value of $\bb{E} \bcdot \bb{B}$ is equivalent to the dot product of the interpolated $\bb{E}$ and the interpolated $\bb{B}$ (see \S3.2.2 of ref.~\cite{lpq09}).}

Our code offers two methods for calculating moments of the ion distribution function. In the full-$f$ method ($\lambda = 1$ in eq.~\ref{eqn:deposit} below), the particles are taken to represent the full distribution function, $f(t, \bb{r}, \bb{v})$. Integral moments of the full distribution function are represented as discrete sums over moments of all the particles. In the $\delta f$ method ($\lambda = 0$ in eq.~\ref{eqn:deposit}), the particles are Lagrangian markers taken to represent the difference $\delta f$ between the full distribution function and a known equilibrium distribution function, $f_0(\bb{r}, \bb{v})$ \cite{pl93,hk94,dk95,bdc97,bjjyk00,cpcu13}. Integral moments of the full distribution function are represented as analytic integrals over $f_0(\bb{r} , \bb{v})$ plus discrete sums over moments of all particles weighted by
\begin{subequations}
\begin{align}
w_p & \equiv \frac{\delta f (t, \bb{x}_p(t) , \bb{v}_p(t) ) }{ f (t, \bb{x}_p(t) , \bb{v}_p(t) ) } \\*
\mbox{} & = 1 - \frac{ f_0 (\bb{x}_p(t) , \bb{v}_p(t) ) }{ f (0, \bb{x}_p(0) , \bb{v}_p(0)) } \quad {\rm along~characteristics.}
\end{align}
\end{subequations}
The ion number density, mean momentum density, and pressure obtained in this fashion are, respectively,
\begin{subequations}\label{eqn:deposit}
\begin{equation}
n(t,\bb{r}) \equiv \int {\rm d}^3 \bb{v}\, f(t, \bb{r}, \bb{v} )  \simeq (1- \lambda) \, n_0(\bb{r}) + \sum_{p=1}^{N_{\rm p}} S( \bb{r} - \bb{x}_p ) \bigl[ \lambda + ( 1 - \lambda ) \, w_p \bigr] ,
\end{equation}
\begin{equation}
n(t,\bb{r}) \, \bb{u}(t,\bb{r})\equiv \int  {\rm d}^3 \bb{v}\, \bb{v} \, f(t, \bb{r}, \bb{v} ) \simeq ( 1 - \lambda ) \, n_0(\bb{r} ) \, \bb{u}_0 (\bb{r} ) + \sum_{p=1}^{N_{\rm p}} \bb{v}_p \,S( \bb{r} - \bb{x}_p )  \bigl[ \lambda + ( 1 - \lambda ) \, w_p \bigr] ,
\end{equation}
\begin{equation}
\msb{P} ( t , \bb{r} ) \equiv \int  {\rm d}^3 \bb{v}\, \bb{v} \bb{v} \, f(t, \bb{r}, \bb{v} ) \simeq  ( 1 - \lambda ) \, \msb{P}_0(\bb{r} ) + \sum_{p=1}^{N_{\rm p}} \bb{v}_p \bb{v}_p \,S( \bb{r} - \bb{x}_p )  \bigl[ \lambda + ( 1 - \lambda ) \, w_p \bigr] .
\end{equation}
\end{subequations}
We emphasize that the $\delta f$ method, when written in this form, is fully nonlinear and so its use need not be restricted to situations in which $\delta f \ll f_0$. However, one must be mindful that this method does not guarantee non-negative values for the total distribution function and so caution should be exercised when using this method with $\delta f \sim f_0$.

Common choices for $f_0$ are the Maxwellian distribution function,
\begin{equation}\label{eqn:maxwellian}
f_{\rm M}(v) \equiv \frac{n_0}{( \pi \beta_0)^{3/2}} \exp \left( - \frac{v^2}{\beta_0} \right), ~\textrm{for which~} n_0(\bb{r}) = n_0, ~\bb{u}_0(\bb{r}) = 0, ~{\rm and}~\msb{P}_0(\bb{r}) = \frac{1}{2} n_0 \beta_0 \msb{I} ;
\end{equation}
and the bi-Maxwellian distribution function,
\begin{eqnarray}\label{eqn:bimaxwellian}
\lefteqn{
f_{\textrm{bi-M}} ( v_{||}, v_\perp ) \equiv \frac{ n_0 }{ \pi^{3/2} \beta^{1/2}_{||,0} \beta_{\perp,0} } \exp\left( - \frac{v^2_{||}}{\beta_{||,0}} - \frac{v^2_\perp}{\beta_{\perp,0}} \right),
}\nonumber\\*&&
\mbox{} \textrm{for which~} n_0(\bb{r}) = n_0,~\bb{u}_0(\bb{r}) = 0,~{\rm and}~\msb{P}_0(\bb{r}) = \frac{1}{2} n_0 \left[ \beta_{\perp,0} \msb{I} - \left( \beta_{\perp,0} - \beta_{||,0} \right) \eb_0 \eb_0 \right] .
\end{eqnarray}
In the latter, $\beta_{\perp,0}$ and $\beta_{||,0}$ are, respectively, the initial plasma beta parameters perpendicular and parallel to the direction of the initial magnetic field, $\eb_0 \equiv \bb{B}_0 / B_0$.

Calculating moments of the distribution function with a finite number of particles $N_{\rm p}$ introduces numerical noise $\propto$$N_{\rm p}^{-1/2}$. While this noise is reduced by a factor $\sim$$| \delta f / f |^2$ in the $\delta f$ method,\footnote{Another approach to reducing noise when solving the hybrid-kinetic equations is to employ a Eulerian grid-based approach \cite[e.g.][]{vtchm07} instead of a Lagrangian PIC method.} both methods are greatly improved by applying a digital filter. We have provided the option of applying up to five filter operations on the number and momentum densities. Each filtering operation is performed with a three-dimensional filtering kernel, which is separable into three one-dimensional filter passes with the filtered value of each cell computed using a three-point binomial filter \citep{bl91}. In Fourier space, this filter is equivalent to a low-pass filter, which eliminates high-frequency numerical noise \citep{buneman93}. Since the filtering kernel can be represented as a convolution in each of the three dimensions, the commutativity of convolutions allows the one-dimensional filter passes to be applied in any order. We have optimized the filter to reduce memory movement and increase cache locality by using vectorization techniques developed by L.~Garrison (private communication) for the electromagnetic relativistic PIC code Tristan-MP \cite{buneman93,spitkovsky05}. We emphasize that digital filtering is not required for code stability in {\it Pegasus}, but rather is used solely to combat finite-number particle noise and thereby provide smoother density and velocity fields.

\subsection{Constrained Transport}\label{sec:CT}

Hybrid-PIC codes generally employ a staggered mesh to facilitate the centering of discretized derivatives and the interpolation of fields to particle locations. However, unless the variables are staggered and the derivatives discretized in specific ways, the divergence-free constraint on the magnetic field will be broken by truncation error \cite{bb80,ramshaw83,eh88}. One way to circumvent this issue is to work with a magnetic vector potential satisfying $\bb{B} = \grad \btimes \bb{A}$ \cite{bcch78,bdc97,bjjyk00}. As long as one's finite-difference representations of the divergence and curl operators cause the numerical representation of $\grad \bcdot \grad \btimes$ to vanish identically, this approach insures maintenance of the solenoidality constraint. 

However, there are several reasons motivating us to work directly with the magnetic field itself. First, the magnetic field occurs explicitly in both the equations of motion for the ion particles and in the expression for the electric field. This would necessitate finite differencing a vector potential several times per time-step at different grid locations. Second, the Hall effect would contribute a term $\propto$$( \grad \btimes \grad \btimes \bb{A} ) \btimes ( \grad \btimes \bb{A} )$ to the right-hand side of the evolution equation for $\bb{A}$, whose second-order--accurate discretization requires a relatively large stencil. There is also the problem of gauge freedom: the equations do not constrain the curl-free component of the vector potential and so it may drift in time. 

The choice to work directly with $\bb{B}$ leaves us with three options. First, we could simply ignore the numerical production of magnetic monopoles and hope that the error introduced is inconsequential compared to the noise generated by the finite number of particles. This is the approach taken in, e.g., Harned's hybrid-PIC algorithm \cite{harned82}, {\it QN3D} \cite{hsa89}, and {\it dHybrid} \cite{gbfs07}. This approach is unsatisfactory, however, particularly in $\delta f$ problems where the particle noise is extremely small and the magnetic-field error could dominate. In addition, breaking magnetic-flux conservation in the shearing box is known to significantly affect the evolution of the magnetorotational instability, a key application of our code. A second approach is to adopt a ``divergence-cleaning" method \cite{ramshaw83}, as used in the {\it A.I.K.E.F.} hybrid-PIC code \cite{msmsgp11}. However, this method would have to be applied three times per time-step (\S\ref{sec:algorithm}), four if working in the shearing sheet, and is therefore unnecessarily expensive.

The third approach, and the one we have ultimately chosen, is commonly referred to as ``constrained transport". Constrained transport \cite{eh88} describes the incorporation of the divergence-free constraint directly into the finite-difference equations. While constrained transport has become the standard of choice in most astrophysical MHD codes \cite{sn92,gs05,gs08,sgths08} and in fully kinetic PIC codes, its adoption by the hybrid-PIC community seems to be much less prevalent. (Notable exceptions are Swift's hybrid-PIC algorithm \cite{swift96} and {\it HYPERS} \cite{ok12}.) For this reason, we offer here a brief summary of the method. 

\begin{figure}
\centering
\includegraphics[width=0.45\textwidth,clip]{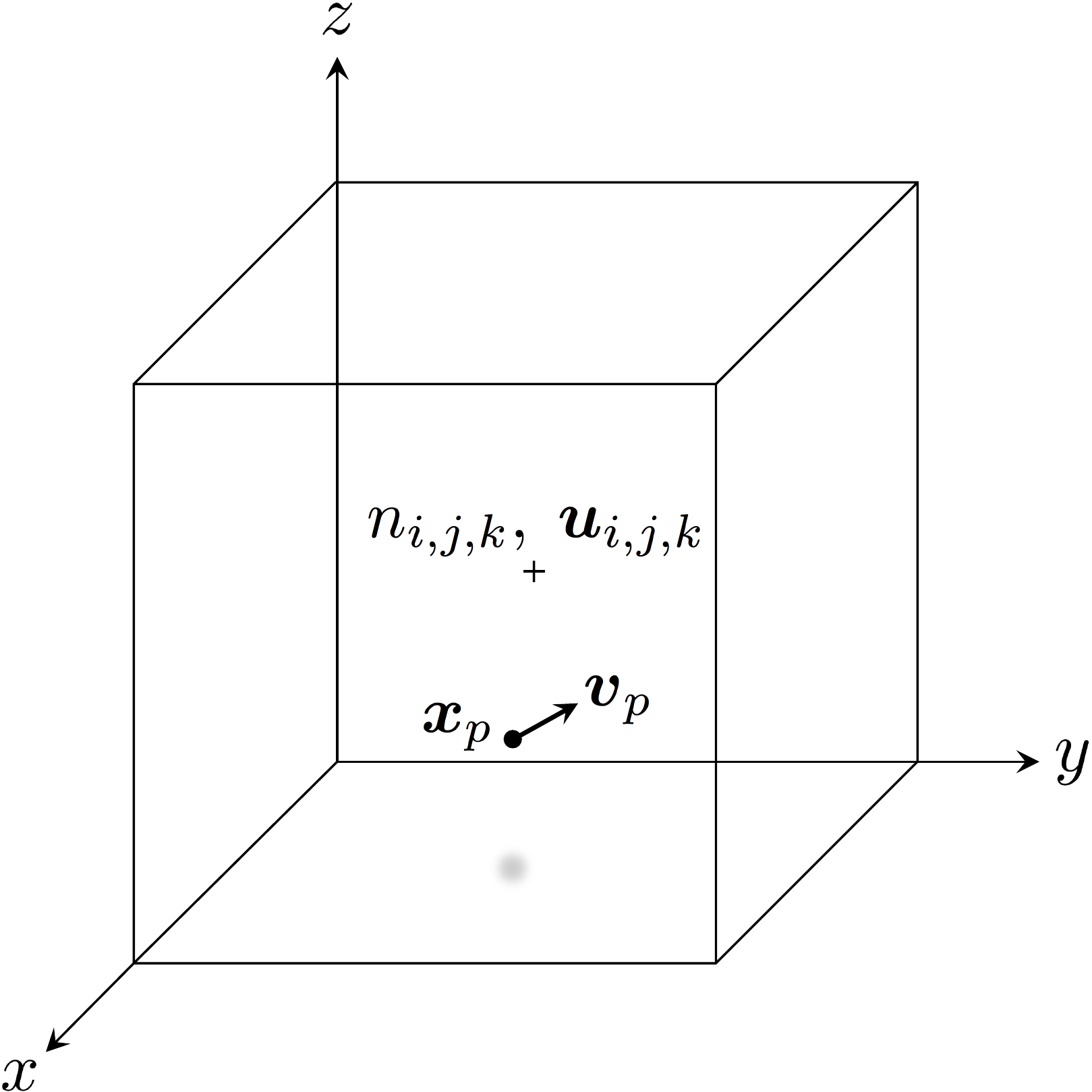}
\quad\quad
\includegraphics[width=0.45\textwidth,clip]{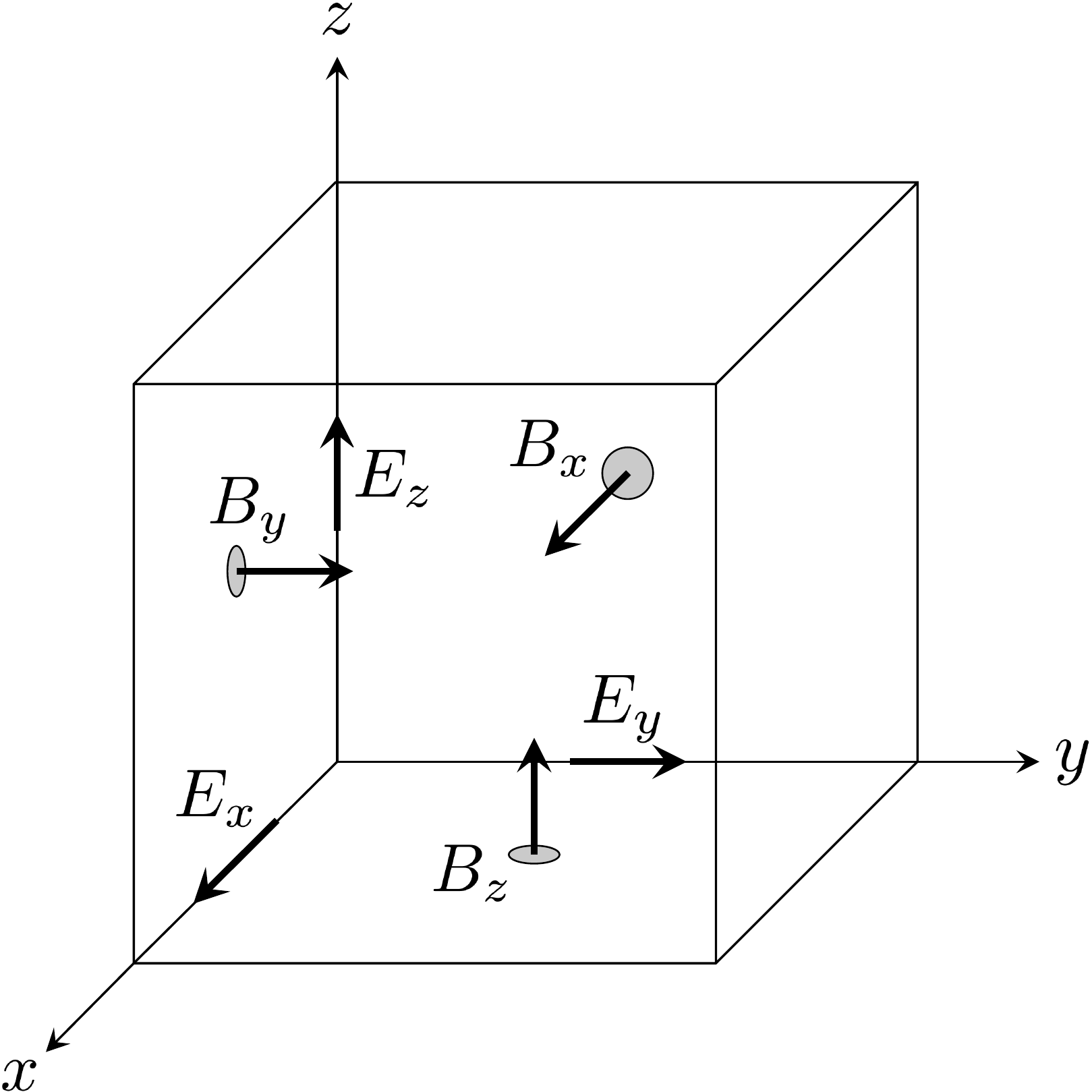}
\caption{The computational mesh used in {\em Pegasus}. ({\it left}) The particles' positions $\bb{x}_p$ and velocities $\bb{v}_p$ ($p=1\dots N_{\rm p}$) are deposited onto the computational mesh (\S \ref{sec:deposit}) using the TSC shape function to obtain the density $n_{i,j,k}$ and velocity $\bb{u}_{i,j,k}$ at the cell center ($i,j,k$). ({\it right}) The magnetic and electric fields are staggered as in the Yee \cite{yee66} mesh, allowing the constrained transport method (\S \ref{sec:CT}) to update the magnetic field in a divergence-free manner.}
\label{fig:grid}
\end{figure}

First, the magnetic-field components are centered on their respective faces of each computational zone and the electric-field components are centered along their respective axes---the so-called Yee \cite{yee66} lattice (see Fig.~\ref{fig:grid}). This arrangement allows each component of $\grad \btimes \bb{E}$ to be naturally centered on the correct magnetic-field component. The second step is to balance the electric-field contributions taken around the cell edges in such as way as to guarantee that the total magnetic flux piercing the surface of a computational zone is always zero (to machine precision):
\begin{subequations}\label{eqn:ct}
\begin{equation}
B^{n+1}_{x; i-1/2,j,k} = B^{n}_{x; i-1/2,j,k} - \frac{h}{\Delta y} \left( E^{n+1/2}_{z; i-1/2, j+1/2, k} - E^{n+1/2}_{z; i-1/2,j-1/2,k} \right) + \frac{h}{\Delta z} \left( E^{n+1/2}_{y; i-1/2, j, k+1/2} - E^{n+1/2}_{y; i-1/2,j,k-1/2} \right) ,
\end{equation}
\begin{equation}
B^{n+1}_{y; i,j-1/2,k} = B^{n}_{y; i,j-1/2,k} - \frac{h}{\Delta z} \left( E^{n+1/2}_{x; i, j-1/2, k+1/2} - E^{n+1/2}_{x; i,j-1/2,k-1/2} \right) + \frac{h}{\Delta x} \left( E^{n+1/2}_{z; i+1/2, j-1/2, k} - E^{n+1/2}_{z; i-1/2,j-1/2,k} \right)  ,
\end{equation}
\begin{equation}
B^{n+1}_{z; i,j,k-1/2} = B^{n}_{z; i,j,k-1/2} - \frac{h}{\Delta x} \left( E^{n+1/2}_{y; i+1/2, j, k-1/2} - E^{n+1/2}_{y; i-1/2,j,k-1/2} \right) + \frac{h}{\Delta y} \left( E^{n+1/2}_{x; i, j+1/2, k-1/2} - E^{n+1/2}_{x; i,j-1/2,k-1/2} \right) .
\end{equation}
\end{subequations}
This is essentially a discrete analogue of Stoke's theorem. Since equations (\ref{eqn:ct}) imply
$$
\left( \grad \bcdot \bb{B} \right)^{n+1}_{i,j,k} = \frac{B^{n+1}_{x;i+1/2,j,k} - B^{n+1}_{x;i-1/2,j,k}}{\Delta x} + \frac{B^{n+1}_{y;i,j+1/2,k} - B^{n+1}_{y;i,j-1/2,k}}{\Delta y} + \frac{B^{n+1}_{z;i,j,k+1/2} - B^{n+1}_{z;i,j,k-1/2}}{\Delta z} = 0 ,
$$
the magnetic field remains divergence-free provided it was so initially. For simulations where a non-uniform magnetic field is a part of the initial state, we utilize a finite-differenced vector potential to insure $\grad \bcdot \bb{B} = 0$ initially to machine precision.

In {\it Pegasus}, the primary description of the magnetic field is taken to be the face-centered values; these are the fields that are evolved using equations (\ref{eqn:ct}). However, cell-centered values for the field are needed for interpolation to the particle positions. At the end of each magnetic-field update, we compute the cell-centered fields as second-order--accurate averages:
\begin{subequations}\label{eqn:bcenter}
\begin{equation}
B_{x;i,j,k} = \frac{1}{2} \left( B_{x;i+1/2,j,k} + B_{x; i-1/2,j,k} \right) , 
\end{equation}
\begin{equation}
B_{y;i,j,k} = \frac{1}{2} \left( B_{y;i,j+1/2,k} + B_{y;i,j-1/2,k} \right) ,
\end{equation}
\begin{equation}
B_{z;i,j,k} = \frac{1}{2} \left( B_{z;i,j,k+1/2} + B_{z;i,j,k-1/2} \right) .
\end{equation}
\end{subequations}
These cell-centered fields only ever serve as reference fields for interpolation to the particle positions. As we have not taken any measures to enforce the divergence-free constraint on the interpolated fields, the cell-centered approximation (\ref{eqn:bcenter}) should not introduce any further errors. Similarly, two sets of electric fields are computed: one at the cell edges and one at the cell centers. The former are used in the constrained transport algorithm, while the latter are used for interpolating the electric field to the positions of the particles.

%
% Figure 1
%

\subsection{Orbital advection and shearing-periodic boundary conditions}\label{sec:fargo}

Splitting the velocity into an ordered background linear shear and a peculiar velocity allowed us to construct an energy-conserving algorithm for particle pushes in the shearing sheet. It also insures that the background shear is never relaxed by noise, an inevitability when a finite number of collisionless particles are entrusted with carrying the full velocity information. In the accretion-disk literature this simplification is referred to as ``orbital advection", an approach that has been employed in both grid-based \cite{masset00,gammie01,jgg08,sg10} and spectral \cite{ur04,ll05,jyk09} hydro- and magnetohydrodynamical codes. Not only does the time-step constraint for stability become independent of the magnitude of the local shear velocity (a non-trivial advantage when considering supersonic shear flows), but also the accuracy of integration is improved by making the truncation error more uniform in radius \cite{jgg08,jyk09}.

For the particle update, orbital advection was implicit in the algorithm described in Section \ref{sec:boris}. The key step is that, after the modified Boris push has been applied, the particle positions are shifted in the azimuthal direction by $-h \sigma \bar{x}$. For the shear advection and stretching of the magnetic field, we employ the second-order--accurate algorithm developed by Stone and Gardiner \cite{sg10} for the MHD Godunov code {\it Athena}. At the end of each time-step, the electric field induced by the shearing motion is integrated along the orbital characteristics a distance $h \sigma x$ upstream of the appropriate cell edges,
\begin{equation}\label{eqn:eshear}
\bb{E}_{\rm shear} \equiv \frac{1}{h} \int_{y}^{y+ h \sigma x} \ey \btimes \bb{B} ~ {\rm d}y ,
\end{equation}
and then is used with constrained transport (eq.~\ref{eqn:ct}) to update the magnetic field.  When the sheared distance $h \sigma x$ is a non-integer number of grid zones, second-order--accurate van Leer reconstruction is used to compute the fractional part of the integral. Using constrained transport guarantees the conservation of magnetic flux to machine precision. We refer the reader to \S5 of ref.~\cite{sg10} for further details.

Non-axisymmetric solutions of the shearing-box equations require special boundary conditions that offset the solutions by the distance the radial edges of the domain have been displaced by the background shear \cite{hgb95}:
\begin{equation}
f(x,y,z) \rightarrow f(x \pm L_x , y \mp \sigma L_x t, z) .
\end{equation}
We implement these shearing-periodic boundary conditions in the same way as in {\it Athena} \cite{sgths08}. Periodic boundary conditions are applied in the radial direction, followed by a conservative remap of all quantities in the ghost cells in the $y$-direction by a displacement $\sigma L_x t$, the distance the boundaries have sheared in time $t$. As this distance is generally not an integer number of cells, second-order--accurate van Leer reconstruction is used to compute the amount of each variable to assign to the remapped cells. While this procedure does not maintain the divergence-free constraint in the ghost zones, these zones are never used for anything other than interpolation. That being said, shearing-box boundary conditions can destroy magnetic-flux conservation if the integral of the electric fields over the two radial faces are not identical after the remap \cite{gz07}. To circumvent this issue, we have followed Stone and Gardiner \cite{sg10} by remapping the azimuthal component of the electric field at each radial face, and using the arithmetic average of $E_y$ from the corresponding grid zone on the opposite face to update the magnetic field in the zones next to the boundary. This procedure conserves the net magnetic flux in the vertical direction to machine precision. We apply a similar procedure to all three components of the current density prior to the calculation of the electric fields, a step inspired by work on ambipolar diffusion in the shearing box (X.~Bai, private communication) finding that errors are otherwise introduced at the shearing boundaries. We refer the reader to \S4 of ref.~\cite{sg10} for further discussion, particularly on how these procedures are implemented when using domain decomposition in parallel computing.

\subsection{The integration algorithm}\label{sec:algorithm}

The difficulty in any hybrid-kinetic code is obtaining an accurate estimate of the time-advanced electric field, which is necessary to perform second-order--accurate updates of the particle positions and velocities (eq.~\ref{eqn:push}) and of the magnetic field (eq.~\ref{eqn:ct}). The difficulty lies in the fact that the electric field does not satisfy an evolutionary equation, which could be leapfrogged with the magnetic field as in a relativistic electromagnetic PIC code \cite{buneman93}, but rather serves as a quasi-neutrality constraint. Since this constraint must be applied at both the integer and half-integer time-steps, all the information required to compute the electric field must be known there as well and so the equations are fundamentally implicit.

The majority of hybrid-PIC codes overcome this difficulty by employing a predictor-corrector method. Often, forward-in-time linear extrapolation is used to obtain a first prediction, using the two previous values of the electric and magnetic fields, which is subsequently iterated to convergence \cite{harned82}. Another class of methods is based upon a moment approach. For example, the ``Current Advance Method" \cite{matthews94} uses a moment equation to estimate the ion current density at the half-step, while subcycling the induction equation to update the magnetic field on fractional time-steps. This method avoids more than one particle push per time-step, a key advantage in the days when computational memory was limited and a prudent programmer avoided unnecessary passes through the particle table.\footnote{The interested reader may consult Appendix A of ref.~\cite{kkvhv04} for a direct comparison of these methods.}

%
% FIgure 2
%
\begin{figure*}
\centering
\includegraphics[width=0.75\textwidth,clip]{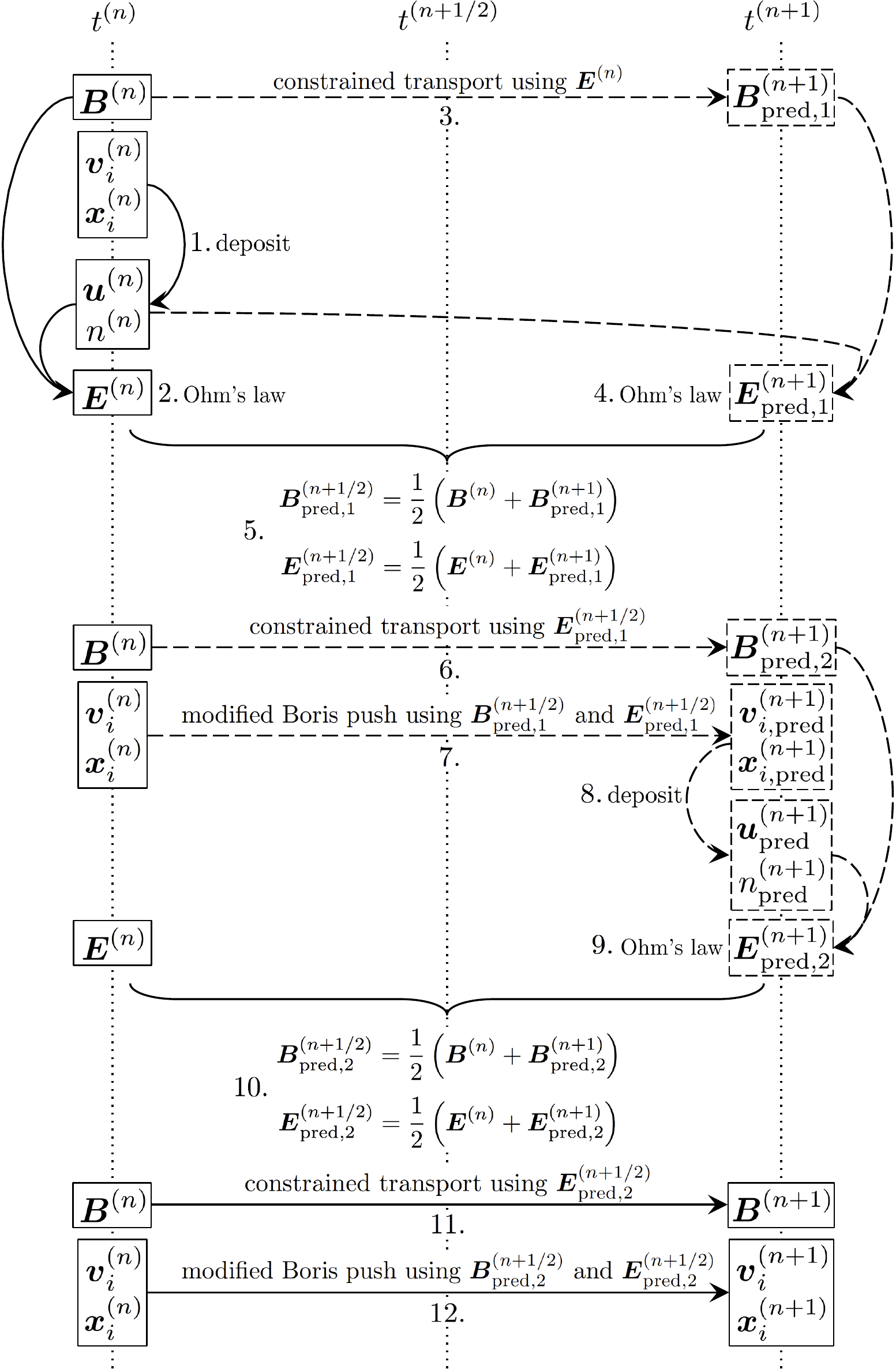}
\caption{Diagram of the integration algorithm used in {\it Pegasus} to update the magnetic field $\bb{B}$, particle velocities $\bb{v}_i$, and particle positions $\bb{x}_i$ from time $t^{(n)}$ to time $t^{(n+1)}$. The steps performed are numbered sequentially 1--12. Solid lines denote tasks that compute accepted (i.e.~permanently stored) values; dashed lines denote temporary steps taken to compute predicted values (subscript ``pred"). See Section \ref{sec:algorithm} for details.}
\label{fig:algorithm}
\end{figure*}

Here we take advantage of modern advances in computational power and memory to construct a predictor-predictor-corrector algorithm that is second-order--accurate in time and can stably propagate whistler waves. In doing so, we have borne three principles in mind. First, particle trajectories should be entrusted with determining time-advanced values of the electric and magnetic fields---not linear extrapolation, nor moment equations. This is accomplished in {\it Pegasus} through a prediction-step particle push to obtain time-advanced values of density and velocity. Secondly, it is straightforward to show by a von Neumann stability analysis that, for a directionally unsplit algorithm, one must employ at least three stages in order to stably propagate whistler waves (see ref.~\cite{kl13}); integration methods with only two stages, such as second-order Runge-Kutta and Lax-Wendroff, are unconditionally unstable without explicit dissipation.\footnote{In fact, second-order Runge-Kutta is unstable for any problem with real eigenfrequencies.} This suggests the adoption of a third-order Runge-Kutta algorithm. However, its use would require us to abandon our energy-conserving particle push (eq.~\ref{eqn:push}). Our compromise is a predictor-corrector-corrector algorithm based an iterative application of Heun's method \cite{gear71} to the induction equation (eq.~\ref{eqn:faraday}). This so-called iterative Crank-Nicholson method is formally second-order--accurate, is stable to whistler-wave propagation [if $h \le n ( \Delta x )^2 / 2 B$], is total-variation-diminishing, and can be easily incorporated into our two-stage predictor-corrector particle push. This brings us to our third principle. In both the magnetic-field and particle integrations, we have sought to construct a definition of the time-centered electromagnetic fields that is as close to time-reversible as possible by always using the trapezoidal rule:\footnote{This is in contrast to the popular integration scheme introduced by Horowitz et al.~for the hybrid-PIC code {\it QN3D} \cite{hsa89}, in which the half-step electric field is computed using time-averaged magnetic fields, i.e.~$\bb{E}^{(n+1/2)} = \bb{E} ( \bb{B}^{(n+1/2)} )$ with $\bb{B}^{(n+1/2)}$ calculated as in equation (\ref{eqn:averaging}). Our scheme instead calculates the half-step electric field by arithmetically averaging the current and time-advanced electric fields (eq.~\ref{eqn:averaging}), each of which is computed using cotemporal magnetic-field values. This makes our scheme total-variation-diminishing in time.}
\begin{equation}\label{eqn:averaging}
\bb{E}^{(n+1/2)} = \frac{1}{2} \left( \bb{E}^{(n)} + \bb{E}^{(n+1)} \right) \quad {\rm and} \quad \bb{B}^{(n+1/2)} = \frac{1}{2} \left( \bb{B}^{(n)} + \bb{B}^{(n+1)} \right) .
\end{equation}
Note that the minimum number of iterations required to render an iterative Crank-Nicholson scheme (conditionally) stable is two (i.e.~there are three stages). Performing more than two iterations does not improve the time-accuracy of the solution beyond second order; in fact, subsequent iterations are not guaranteed to yield stable integration \cite{teukolsky00}. While the amplification factor does approach unity as the number of iterations are increased, it does so non-monotonically. For example, taking four or five iterations yields an unconditionally unstable algorithm for both linear advection and whistler-wave propagation. This is counter-intuitive, as an infinite number of iterations corresponds to the fully implicit (and unconditionally stable) Crank-Nicholson method. This calls into question predictor-corrector methods that iterate until a desired accuracy is achieved without regard for whether a given iteration yields a stable solution.

Our predictor-predictor-corrector algorithm is diagrammed in Figure \ref{fig:algorithm}. At the beginning of each time-step, the particles are deposited onto the active cells of the grid and the number density and momentum density of the ions are calculated at the cell centers using equation \ref{eqn:deposit} (step 1 in Fig.~\ref{fig:algorithm}). Digital filtering is performed on these quantities a user-specified number of times to reduce high-frequency noise. The ghost cells are then filled by copying information from the active cells while respecting the user-specified boundary conditions (\S\ref{sec:bcs}). If $\sigma \ne 0$, the information in the ghost cells is remapped in the $y$-direction by a displacement $\sigma L_x t$, the distance the boundaries have sheared in a time $t$ (\S\ref{sec:fargo}). The information in both the active and ghost cells is used in equation (\ref{eqn:efield}) to obtain a second-order--accurate representation of the electric field $\bb{E}^{(n)}$ at both the centers and edges of every cell in the computational domain (step 2). With all the necessary information at $t^{(n)}$ in hand, we proceed to calculate the time-centered electric field as follows. 

First, we use constrained transport (eq.~\ref{eqn:ct}) with $\bb{E}^{(n)}$ to obtain a prediction for the magnetic field $\bb{B}^{(n+1)}_{\rm pred,1}$ at $t^{n+1}$ (step 3). This magnetic field is fed into equation (\ref{eqn:efield}) to obtain an estimate of the electric field $\bb{E}^{(n+1)}_{\rm pred,1}$ at $t^{n+1}$ (step 4). Note that this prediction assumes fixed density and velocity. Next, these values of the electric and magnetic fields are fed into equation (\ref{eqn:averaging}) to compute $\bb{E}^{(n+1/2)}_{\rm pred,1}$ and $\bb{B}^{(n+1/2)}_{\rm pred,1}$, respectively (step 5). These estimated time-centered fields are then used in equations (\ref{eqn:push}) and (\ref{eqn:ct}) to obtain temporary copies of the magnetic field $\bb{B}^{(n+1)}_{\rm pred,2}$ (step 6) and particle velocities $\bb{v}^{(n+1)}_{i{\rm ,pred}}$ and positions $\bb{x}^{(n+1)}_{i{\rm ,pred}}$ (step 7) at $t^{(n+1)}$. In order to avoid two passes through the particle table, each particle copy is deposited onto the grid immediately following its update and the number and momentum densities of the ions are calculated at the cell centers (step 8).

At this stage of the algorithm, we have predicted values for the number density, momentum density, and magnetic field at $t^{n+1}$. As before, digital filtering is performed on the number density and momentum density a user-specified number of times. The ghost zones are filled by copying information from the active cells into the ghost cells, while respecting the user-specified boundary conditions. If $\sigma \ne 0$, the information in the ghost cells is remapped in the $y$-direction by a displacement $\sigma L_x t$. These values are fed into equation (\ref{eqn:efield}) to obtain an estimate of the electric field $\bb{E}^{(n+1)}_{\rm pred,2}$ at $t^{n+1}$ (step 9). The magnetic and electric fields at $t^{(n+1/2)}$ are once again found by arithmetic averaging (eq.~\ref{eqn:averaging}; step 10).

Once values for the electric and magnetic fields at $t^{(n+1/2)}$ are known, the magnetic field (step 11) and particle positions and velocities (step 12) can finally be updated from $t^{(n)}$ to $t^{(n+1)}$ using equations (\ref{eqn:ct}) and (\ref{eqn:push}), respectively. If $\sigma \ne 0$, orbital advection is used to update the particle positions and the magnetic field (\S\ref{sec:fargo}). Boundary conditions are applied to the magnetic field, crossing particles are shared between processors, and additional particles are injected if requested (\S\ref{sec:bcs}). If desired, I/O functions are performed. Finally, a test is performed to verify that the time-step is shorter than that allowed by the Courant condition
\[
h \le \mc{C} \times {\rm min} \left( \frac{\Delta x}{v_{\rm max}} , \frac{2\pi}{B} \right)\, , \quad {\rm where} \quad v_{\rm max} = {\rm max} \left( \left| v_p \right|, \left| u \right| + v_{\rm A} + \frac{2 B}{n \Delta x} \right)
\]
and $\mc{C}$ is the Courant number. If successful, the time is incremented by $h$.

\subsection{Boundary conditions}\label{sec:bcs}

In addition to shearing-periodic boundary conditions, {\it Pegasus} offers reflecting, conducting, outflow, inflow, or periodic boundary conditions on any domain boundary. User-defined boundary conditions may be imposed on any domain boundary via a modular routine provided in the user's problem generator. Boundary conditions are generally set at the end of each time-step, after the integration has been performed and before I/O functions are called. However, the boundary conditions must also be set during the integration when the particles are deposited onto the grid to compute the zeroth and first moments (twice per time-step; see Section \ref{sec:algorithm}). This is because the TSC shape function deposits fractions of these moments into the ghost zones, from which they must be remapped back into the appropriate active zones while respecting the boundary conditions. For example, the portion of a momentum deposit that overlaps a reflecting boundary is removed from the ghost zones and re-deposited back into the neighboring active zones with its sign reversed. When shearing-periodic boundary conditions are used, any portion overlapping a radial boundary is remapped to the opposite radial boundary and shifted in the $y$-direction (as in Section \ref{sec:fargo}) before being re-deposited. This procedure requires two additional MPI calls per time-step, adding to the computational cost. However, we found this technique to be less computationally expensive than adding additional ``ghost" particles every time-step, which, when deposited, would naturally fill the ghost zones with the appropriate moments.

Particle injectors are available on any domain boundary. One simply activates the desired injectors via a boundary-condition flag and sets the injection speed. The injector then places enough active particles (drawn from a user-specified distribution function) into the ghost zones so that, when integrated and naturally evolved onto the active grid, the desired number of particles per cell results on average in the cells neighboring the injection boundary.

\subsection{Implementation and performance}\label{sec:implementation}

In assimilating the above algorithms to create {\it Pegasus}, we have adopted the well-documented and thoroughly tested {\it Athena} architecture \cite{sgths08}. We believe the similarity in these codes will facilitate ease of use within an astrophysical numerical community already acquainted with the widely used {\it Athena} MHD code. As such, {\it Pegasus} is written in C, is modular in design, strictly adheres to ANSI standards, and offers the user simple control of physics and runtime options. Regarding the latter, physics and algorithm options are set at compile time using a configure script generated by the {\tt autoconf} toolkit. All problem-specific code is contained in a single file, with functionality provided that makes it easy to add new boundary conditions, new distribution functions, new I/O quantities, and even new source terms in the equations.

We have also adopted a similar philosophy regarding code optimization as that used in {\it Athena}: since aggressive optimization often comes at the cost of clarity and adaptability in the code, we have limited optimization to the basic concepts guided by the rules of data locality and vectorization. For example, cache access is optimized by defining all grid variables within a cell as a data structure and then creating three-dimensional arrays of this structure. Similarly, all information for each particle [i.e.~$\bb{x}_p$, $\bb{v}_p$, particle properties, particle identification number $p$, initial host processor number, and $f (0, \bb{x}_p(0) , \bb{v}_p(0))$ if the $\delta f$ method is used] are stored as a data structure; an $N_{\rm p}$-dimensional array of this structure is then constructed as the particle table. These measures ensure that values for each variable associated with a given cell or a given particle are contiguous in memory. We have further implemented a quicksort algorithm to periodically re-order the particles to achieve high data localization.

{\it Pegasus} is written to run either as a serial code on one CPU or in parallel using domain decomposition through MPI calls. As in {\it Athena}, the sequential exchange of boundary conditions in the $x$-, $y$-, and $z$-directions across processors eliminates the need for extra MPI calls to swap values across diagonal domains at the corners of the grid. Since all the magnetic-field information required by the algorithm is set in the ghost zones by a single MPI call at the start of the time-step, no further communication is required during the algorithm to exchange electric- and magnetic-field information. The only communication that occurs during the integration itself is that required to remap onto the active cells those portions of the deposited quantities that lie outside of the active grid and set the boundary conditions on the number and momentum densities (\S\ref{sec:bcs}). This occurs twice per time-step. Additional MPI calls are needed in the shearing box to perform the orbital advection of the magnetic field by remapping (\S\ref{sec:fargo}). Finally, at the end of each time-step, any particles that have crossed processor boundaries are shared between processors via MPI calls.

\begin{figure}
\centering
\includegraphics{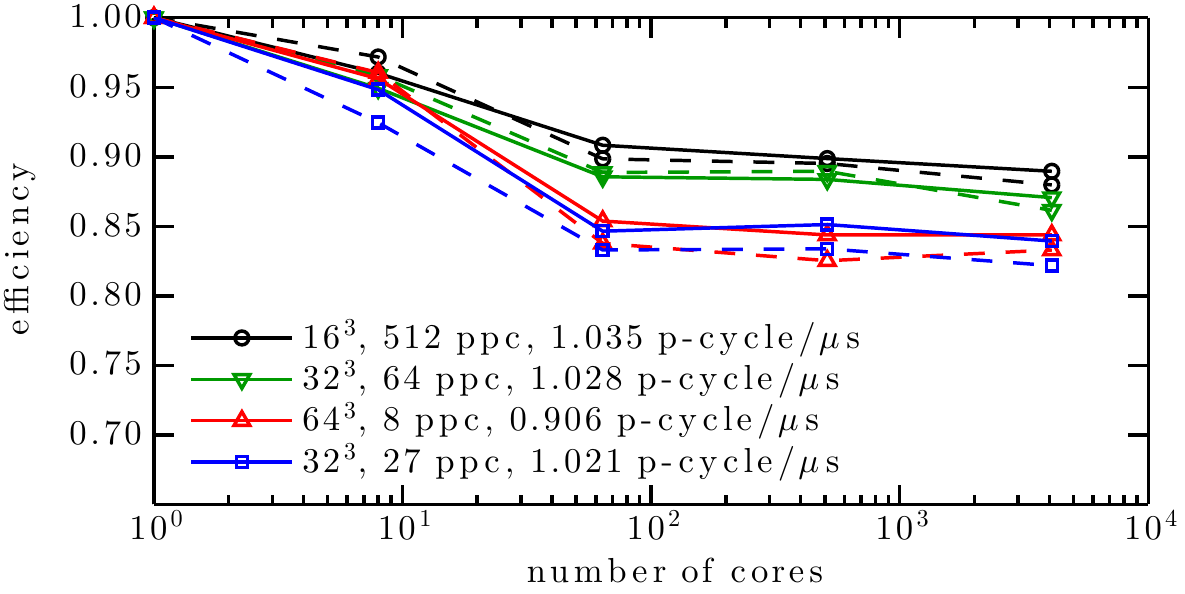}
\caption{Weak scaling of {\em Pegasus} on Stampede for a three-dimensional linear wave problem. The number of cells per core (e.g.~$16^3$) is indicated, as are the number of particles per cell (``ppc"), and the single-core performance in particle-cycles per microsecond. Dashed (solid) lines correspond to one (zero) filter pass(es) per particle deposit. Note that the first three cases listed all share an equal number of particles per core, and that there are 16 cores per node on Stampede.}
\label{fig:scaling}
\end{figure}

A useful performance measure in PIC codes is the number of particles updated per CPU microsecond. This number depends on many factors, including the size of the grid, the number of particles per cell, the processor speed, the number of filter passes used, and the boundary conditions. That being said, we have found that {\it Pegasus} averages $\sim$$1$ particle-cycle per CPU-$\mu$s on modern machines (e.g.~on Stampede at the Texas Advanced Computing Center, an Infiniband network of Dell PowerEdge C8220X nodes with dual 8-core Intel Xeon E5-2680 processors). Since the ratio of computational work to data communicated is large in {\it Pegasus} due to the expense of pushing (potentially millions of) particles per core twice per time-step, the code scales very well to thousands of cores. Figure \ref{fig:scaling} demonstrates weak scaling on Stampede for a three-dimensional linear wave problem (\S\ref{sec:whistler}) using grids with: (black circles) $16^3$ cells per core and 512 particles per cell (``ppc"), (green inverted triangles) $32^3$ and 64 ppc, (red triangles) $64^3$ and 8 ppc, and (blue squares) $32^3$ and 27 ppc. The first three cases all involve 2,097,152 particles per core, while the final case involves 884,736 particles per core. Dashed (solid) lines correspond to one (zero) filter pass(es) per particle deposit. On Stampede, there are 16 cores per node; once these are filled the efficiency flattens out at $>$$80\%$ (relative to one node) all the way out to 4096 cores ($N_{\rm p} \sim 8.5 \times 10^9$), indicating excellent weak scaling.

There are, however, some differences worth noting. The most advantageous configuration, not only in terms of scaling (efficiency $\sim$$90\%$) but also in terms of performance (1.035 particle-cycles per microsecond on a single core), is the $16^3$--512 case. This indicates that it is best to have a high particle-to-cell ratio on each core. Both the single-core performance and the efficiency go down when fewer particles per cell are used (at a fixed number of particles per core). Figure \ref{fig:scaling} also demonstrates that, for a fixed number of cells per core, the efficiency is decreased when fewer particles per cell are used. Since particle pushes are embarrassingly parallel, the overall efficiency is increased when there are more particles per core.

%
% Figure 3
%

%
% Tests
%
\section{Tests}\label{sec:tests}

In this section we present a selection of tests that we have found useful in the development of {\it Pegasus}. These tests are by no means exhaustive, but rather have been chosen to demonstrate the fidelity and versatility of the method. Some of these tests (e.g.~magnetorotational instability, firehose and mirror instabilities) serve as launching points for further research, and we defer a detailed study of their scientific results to dedicated publications already in progress.

\subsection{Particle motion}\label{sec:orbit}

%
% Figure 4
%
\begin{figure}
\centering
\includegraphics{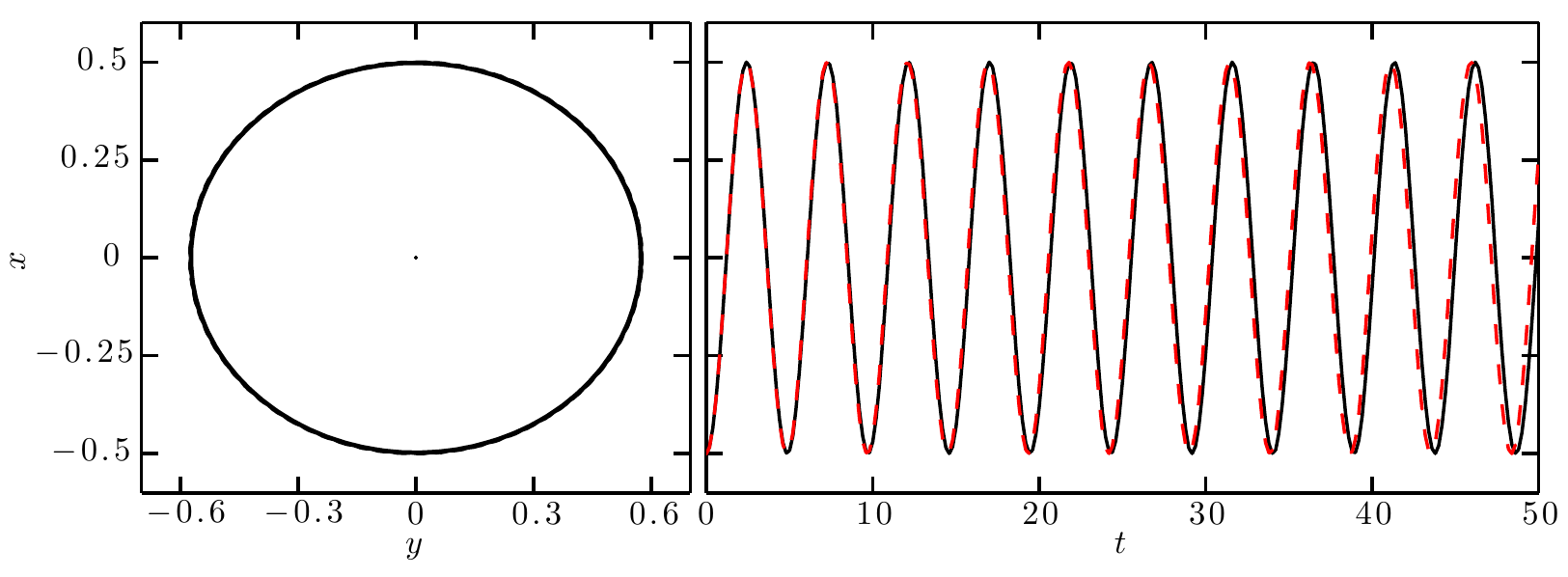}
\caption{Test of particle motion in the shearing sheet with a uniform magnetic field $\bb{B} = \ez$. Shown are ({\it left}) the elliptical particle orbit in the $x$-$y$ plane and ({\it right}) the $x$-coordinate of the particle as a function of time $t$. The dashed red line traces the analytic solution. See Section \ref{sec:orbit} for details.}
\label{fig:orbit}
\end{figure}

We begin our tests by examining the conservation properties of our particle integrator (\S\ref{sec:boris}). To do so, we follow the motion of a charged test particle in the presence of a uniform magnetic field, $\bb{B} = \ez$, embedded in a shearing sheet with $\sigma = (3/2) \varpi$. In a frame co-moving with the gyro-center of the particle,
\begin{equation}
\bb{v}_{\textrm{gyro-center}} = \bigl[ y(0) \ex - x(0) \ey \bigr] \left( \frac{ \omega^2 }{1 + 2 \varpi } \right)  \quad \textrm{where} \quad \omega^2 \equiv ( 1 + 2 \varpi ) ( 1 + 2 \varpi - \sigma ) ,
\end{equation}
the particle should evolve according to the parametric equations
\begin{equation}
x(t) = x(0) \cos \omega t + y(0) \sin \omega t  \left( \frac{ \omega }{ 1 + 2 \varpi } \right) , \quad y(t) = y(0) \cos \omega t - x(0) \sin \omega t \left( \frac{\omega}{1+2\varpi} \right) ,
\end{equation}
while maintaining a constant energy
\begin{equation}
U(t) = U(0) = \frac{1}{2} v^2 - \frac{1}{2} v^2_x  \left( \frac{ \sigma }{ 1 + 2 \varpi } \right) .
\end{equation}
In Figure \ref{fig:orbit}, we plot the numerical solution in the $x$-$y$ and $x$-$t$ planes for $x(0) = -1/2$, $y(0) = 0$, and $\varpi = 1/4$; the time-step $h = 0.2$. The numerical solution conserves energy to machine precision and the elliptical particle orbit is closed. The truncation error manifests as a phase shift relative to the analytic solution (denoted by the red dashed line); this error decreases as $h^2$. In typical numerical simulations, the chosen time-step is much smaller than $h = 0.2$ and so the phase error is often negligible.

\subsection{Landau damping}\label{sec:landau}

%
% Figure 5
%
\begin{figure}
\centering
\includegraphics{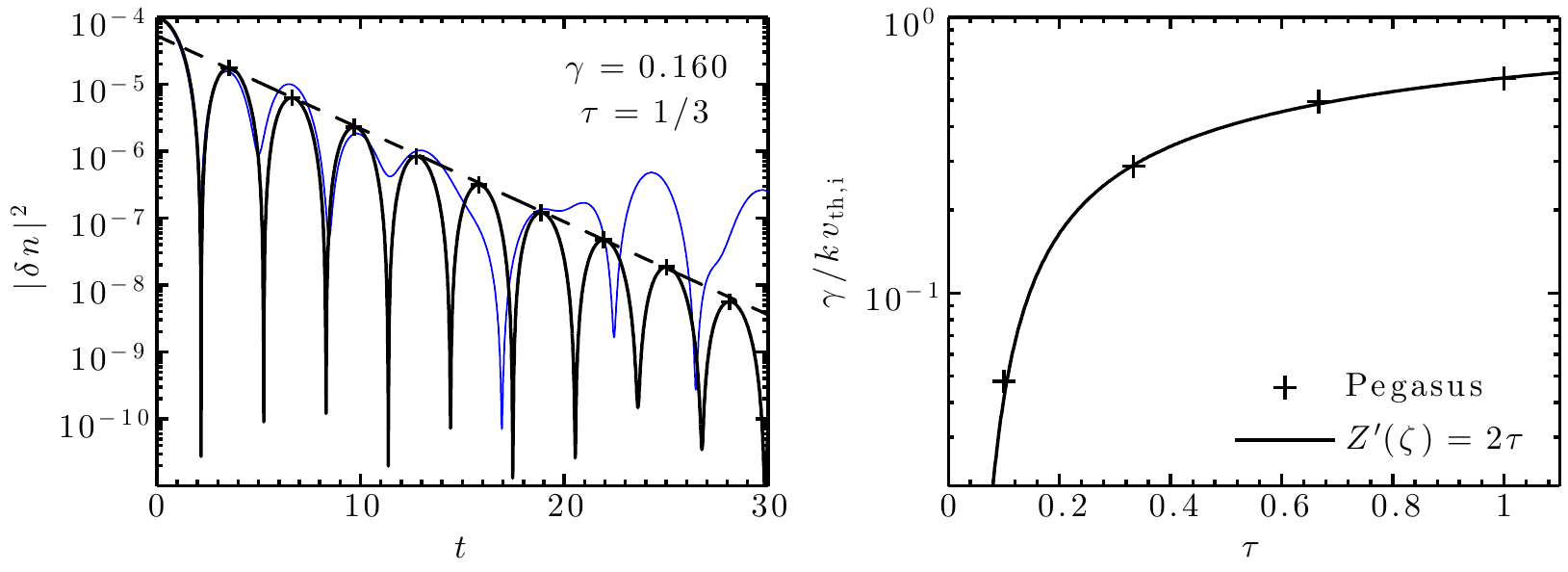}
\caption{Landau damping test. ({\it left}) The numerical solution obtained using the $\delta f$ method (solid black line) for $\tau = 1/3$ accurately follows the expected rate of decay ($\gamma = 0.160$; dashed line) to very low amplitudes. The numerical solution obtained using the full-$f$ method (thin blue line) is shown for comparison. ({\it right}) Numerically determined damping rates (plus signs) for $\tau = 0.1$, $1/3$, $2/3$, and $1$ overlaid on the theoretical curve (solid line). See Section \ref{sec:landau} for details.}
\label{fig:landau}
\end{figure}

Next we test both the $\delta f$ method (\S\ref{sec:deposit}) and the general integration algorithm (\S\ref{sec:algorithm}) by simulating the decay of a small-amplitude ion acoustic wave by Landau damping. In hybrid-kinetics, ion acoustic waves satisfy the dispersion relation ${\rm d}Z(\zeta)/{\rm d}\zeta = 2\tau$, where $Z(\zeta)$ is the plasma dispersion function, and their dimensionless phase velocity $\zeta \equiv (\omega - \imag \gamma) / k v_{\rm th,i}$ exhibits both real and imaginary parts. We seek to numerically recover this dispersion relation by initializing $2.4 \times 10^6$ particles on a one-dimensional periodic grid of spacing $\Delta x = 0.33$ with a Maxwellian velocity distribution (eq.~\ref{eqn:maxwellian}) times a perturbative factor $( 1 + \alpha \cos k x )$. We set the perturbation amplitude $\alpha = 0.01$ and the wavenumber $k = \pi / 8$. In the left panel of Figure \ref{fig:landau} we plot the squared amplitude of the density fluctuation as a function of time for $\tau = 1/3$. The numerical solution (solid black line) accurately follows the expected rate of decay ($\gamma = 0.160$; dashed line) to very low amplitudes. Comparing this numerical solution with one obtained using the full-$f$ method (thin blue line) clearly demonstrates the advantages of reducing particle noise via the $\delta f$ method. In the right panel, we give the numerically determined damping rates (plus signs) for $\tau = 0.1$, $1/3$, $2/3$, and $1$ overlaid on the theoretical curve (solid line). The agreement between the theoretical and numerical results is excellent.

\subsection{Alfv\'{e}n and Whistler waves}\label{sec:whistler}

%
% FIgure 6
%
\begin{figure}
\centering
\includegraphics{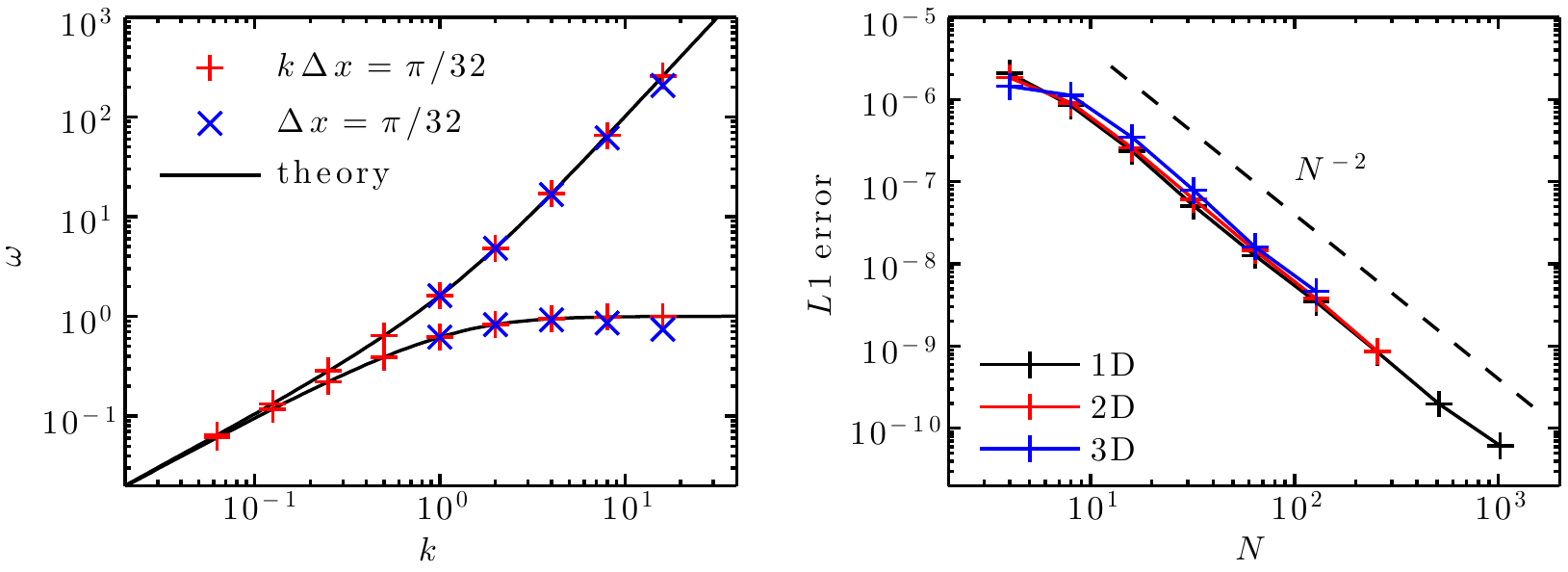}
\caption{Alfv\'{e}n and whistler wave test. ({\it left}) Numerical (plus signs, crosses) and analytical (solid lines) dispersion relations. Circularly polarized Alfv\'{e}n waves ($\omega = k$) appear at low wavenumbers $k\ll 1$, while whistler waves ($\omega = k^2$) manifest at $k \gg 1$. ({\it right}) $L1$ error demonstrating second-order convergence in one, two, and three spatial dimensions. See Section \ref{sec:whistler} for details.}
\label{fig:whistler}
\end{figure}

Whistler waves constitute the high-wavenumber ($k \gg 1$) limit of the Hall-MHD dispersion relation $\omega^2 = k^2 ( 1 \pm \omega)$, while circularly polarized Alfv\'{e}n waves manifest at low wavenumbers $k \ll 1$. The ability to stably and accurately propagate Alfv\'{e}n and whistler waves across the computational grid should be a requirement of any hybrid-PIC code. We tested ours by computing the numerical dispersion relation (Figure \ref{fig:whistler}) for left- and right-handed waves and comparing with the analytical result (solid lines). We chose to work with cold electrons and ions ($\tau^{-1} = \beta = 0$) in order to isolate the propagation of magnetic disturbances and prevent parametric decay into Landau-damped sound waves. The numerical dispersion relation was calculated using two techniques, each using 64 particles per cell and a running time of 4 wave periods. First, we fixed the grid resolution at 64 cells per wavelength ($k \Delta x = \pi / 32$; red plus signs); the agreement is very good for wavelengths up to and beyond the ion skin depth. Next, we fixed the grid resolution at $\Delta x = \pi / 32$ (blue crosses) and increased the wavenumber until the wavelength $\lambda = 4 \Delta x$ (i.e.~one wavelength is resolved by just four cells). Even at this poor resolution, the error in the wave frequency is very small. It is important to note that $\eta = 0$ in this test; waves above and below the ion skin depth propagate stably without the need for explicit magnetic diffusion. In the right panel of Figure \ref{fig:whistler} we plot the $L1$ error as a function of the number of grid cells per wavelength $N$ using $k_{||} \sim 1$, for which the inductive and Hall terms are of comparable magnitude. This exercise is performed in one dimension ($k_{||} = 1$; $N$ cells), two dimensions ($k_{||} = \sqrt{5}/2$; $N \times 2N$ cells), and three dimensions ($k_{||} = 1.5$; $N\times N\times 2N$ cells), with the waveform propagating at an oblique angle to the grid. Second-order convergence is achieved in each case up to the highest resolution tested.

\subsection{Firehose instability}\label{sec:firehose}

%
% Figure 7
%
\begin{figure}
\centering
\includegraphics[clip]{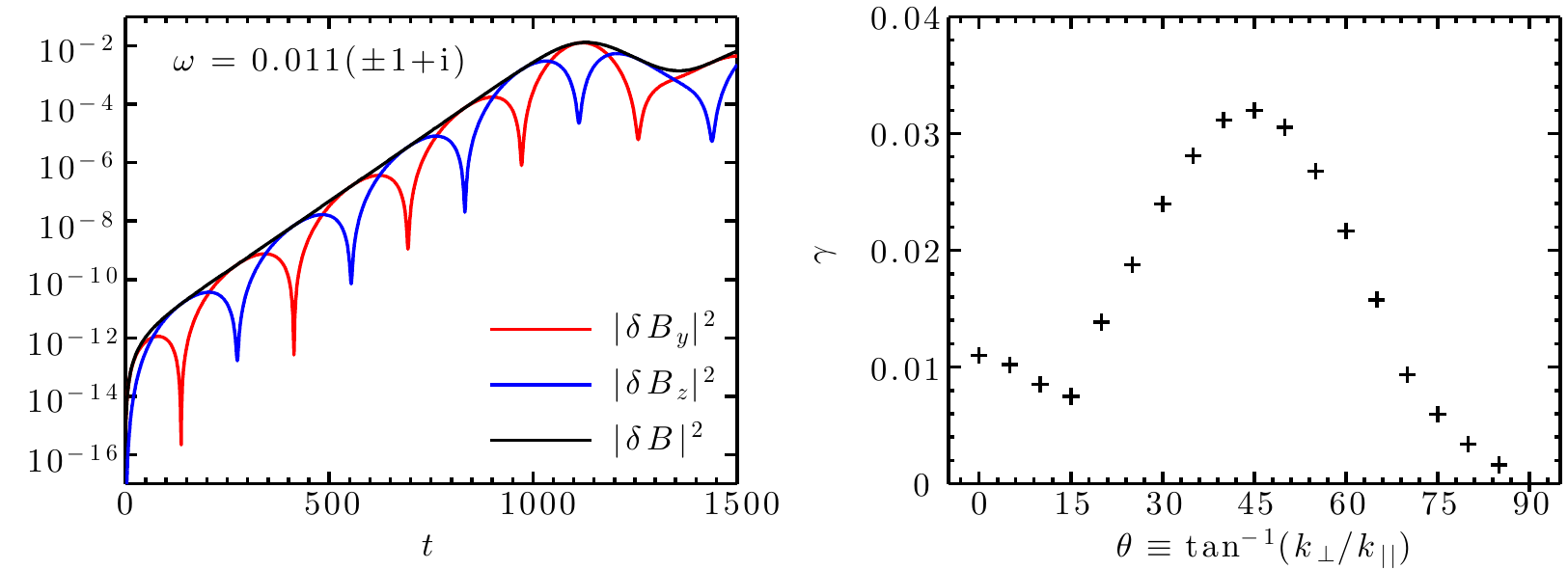}
\caption{Firehose instability test using the $\delta f$ method. ({\it left)} The parallel-propagating firehose instability ($k_\perp = 0$) exhibits a (complex) frequency $\omega = 0.011\,(\pm 1 + \imag)$. ({\it right}) Growth rate $\gamma$ as a function of wavenumber obliquity, $\theta \equiv \tan^{-1} ( k_\perp / k_{||} )$. Alfv\'{e}n-wave--polarized fluctuations occur for $\theta \lesssim 14^\circ$ and are overstable; slow-mode--polarized fluctuations occur at larger $\theta$, are purely growing, and have a peak growth rate $\gamma_{\rm max} = 0.032$ at $\theta_{\rm max} = 45^\circ$. See Section \ref{sec:firehose} for details.}
\label{fig:firehose}
\end{figure}

Many space and astrophysical plasmas exhibit or are very likely to exhibit non-Maxwellian distribution functions with velocity-space anisotropies relative to the local magnetic-field direction. If the resulting pressure anisotropy $p_\perp - p_{||} < -B^2 / 4\pi$, Alfv\'{e}nic perturbations are subject to the firehose instability. While this instability is obtainable in Braginskii-MHD \cite{braginskii65}, its peak wavenumber and growth rate are set by finite-Larmor-radius effects that are captured by evolving kinetic ions. Here we test the ability of our code to accurately produce the firehose instability from an initially bi-Maxwellian distribution function $f_0 = f_{\textrm{bi-M}}$ (see eq.~\ref{eqn:bimaxwellian}) with $\beta_0 = 300 / \pi$ and $\beta_{\perp,0} - \beta_{||,0} = 3$. The instability parameter $\Lambda \equiv ( \beta_{||,0} - \beta_{\perp,0} - 2 ) / \beta_0 = \pi / 300 \simeq 0.011$. In all of the firehose tests, the $\delta f$ method is used and $\eta = 0$.

First, we investigate the parallel-propagating firehose instability \cite{ks67,dv68} by aligning $\bb{B}_0$ with the $x$-direction of a one-dimensional periodic grid with 480 cells and 2048 particles per cell ($N_{\rm p} \sim 10^6$). Assuming isothermal electrons with $\tau = 1$, linear analysis yields a frequency $\omega_{\rm max} = ( \pm 1 + \imag ) \,\Lambda$ and a parallel wavenumber $k_{||{\rm ,max}} \rho_{\rm i,0} = 2 \Lambda^{1/2}$ at maximum growth. The latter yields a parallel wavelength $\lambda_{||{\rm ,max}} = 300$, which we set as the length of the box. The grid spacing is then $\Delta x = 0.625$. In the left panel of Figure \ref{fig:firehose}, we show the evolution of the energy in various components of the perturbed magnetic field. The fastest-growing mode emerges from white noise of amplitude $10^{-6}$ and evolves with (complex) frequency $\omega = 0.011 ( \pm 1 + \imag )$, in agreement with the analytically predicted value. The mode saturates with an amplitude $| \delta \bb{B} |^2 \sim \Lambda \sim 0.01$, in agreement with quasi-linear theory \cite{ss64,sckrh08}. 

In the right panel, we show numerically computed growth rates for the firehose instability as a function of wavenumber obliquity, $\theta \equiv \tan^{-1} ( k_\perp / k_{||} )$; parallel propagation corresponds to $\theta = 0$. For this test, a box of length $600$ comprised of $960$ cells ($\Delta x = 0. 625$) was oriented along the wavenumber direction and the magnetic field was rotated so that $B_y / B_x = k_\perp / k_{||} = \tan\theta$. At small angles ($\theta \lesssim 14^\circ$), the fluctuations are primarily Alfv\'{e}nically polarized and exhibit a complex frequency (i.e.~they are overstable). At larger angles, the fluctuations are slow-mode polarized, are purely growing, and have a peak growth rate $\gamma_{\rm max} = 0.032 \sim \Lambda^{3/4}$ at $\theta = 45^\circ$. The latter is in agreement with linear theory \cite[][A.~Schekochihin, unpublished; A.~Schekochihin and M.~Kunz, in preparation]{ywa93,hm00}.

%
% Figure 8
%
\begin{figure}
\centering
\includegraphics[clip]{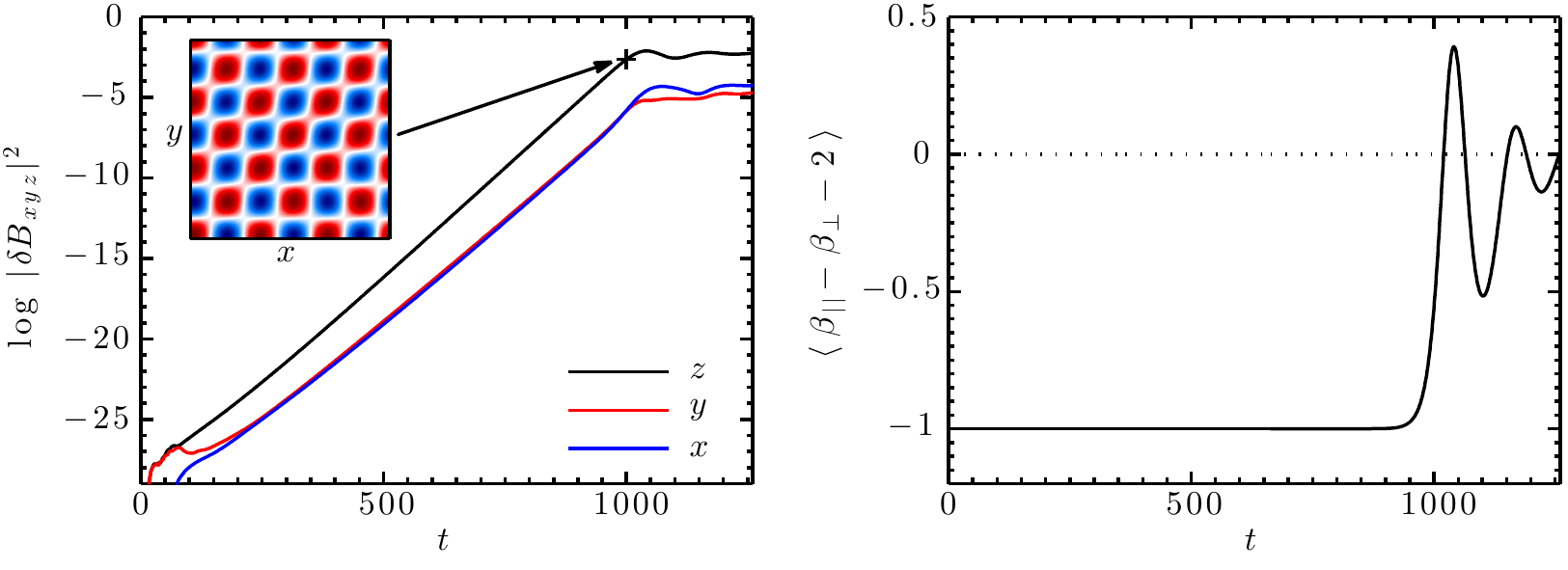}
\caption{Oblique firehose instability test using the $\delta f$ method. ({\it left}) The magnetic fluctuation amplitudes exhibit exponential growth over more than 12 orders of magnitude, during which the $z$-component (whose spatial structure at $t = 10^3$ is shown in the inset) is dominant. ({\it right}) The box-averaged distance to marginal stability, $\langle \beta_{||} - \beta_\perp - 2 \rangle$. See Section \ref{sec:firehose} for details.}
\label{fig:firehose2d}
\end{figure}

Finally, we consider the firehose instability on a two-dimensional periodic grid of $768^2$ cells and 512 particles per cell ($N_{\rm p} \sim 3 \times 10^8$). The box size is $480 \times 480$, yielding a grid spacing $\Delta x = \Delta y = 0.625$. As before, the guide magnetic field is oriented in the $x$-direction; however, the two-dimensional setup allows the system to seek out the fastest-growing oblique mode. Linear theory and the numerical results in Figure \ref{fig:firehose} both show a faster growth rate for oblique modes, and we anticipate that the parallel-propagating mode investigated in the one-dimensional case will be sub-dominant. This is indeed what we find in Figure \ref{fig:firehose2d}, which shows that the perturbed magnetic field ({\it left}) grows purely exponentially at a rate $\gamma \simeq 0.032$. The inset shows the spatial structure of $B_z$ at $t = 10^3$; the emergent mode has $k_\perp \sim k_{||} \sim 0.4 \rho^{-1}_{\rm i}$. Similarly to the parallel-propagating firehose instability, the oblique mode saturates with $| \delta \bb{B} |^2 \sim \Lambda \sim 0.01$, as the distance to marginal stability ({\it right}) tends towards zero. These tests demonstrate the ability of our code to accurately solve for finite-Larmor-radius effects in non-Maxwellian backgrounds.

\subsection{Non-relativistic shock}\label{sec:shock}

%
% Figure 9
%
\begin{figure}
\centering
\includegraphics[clip]{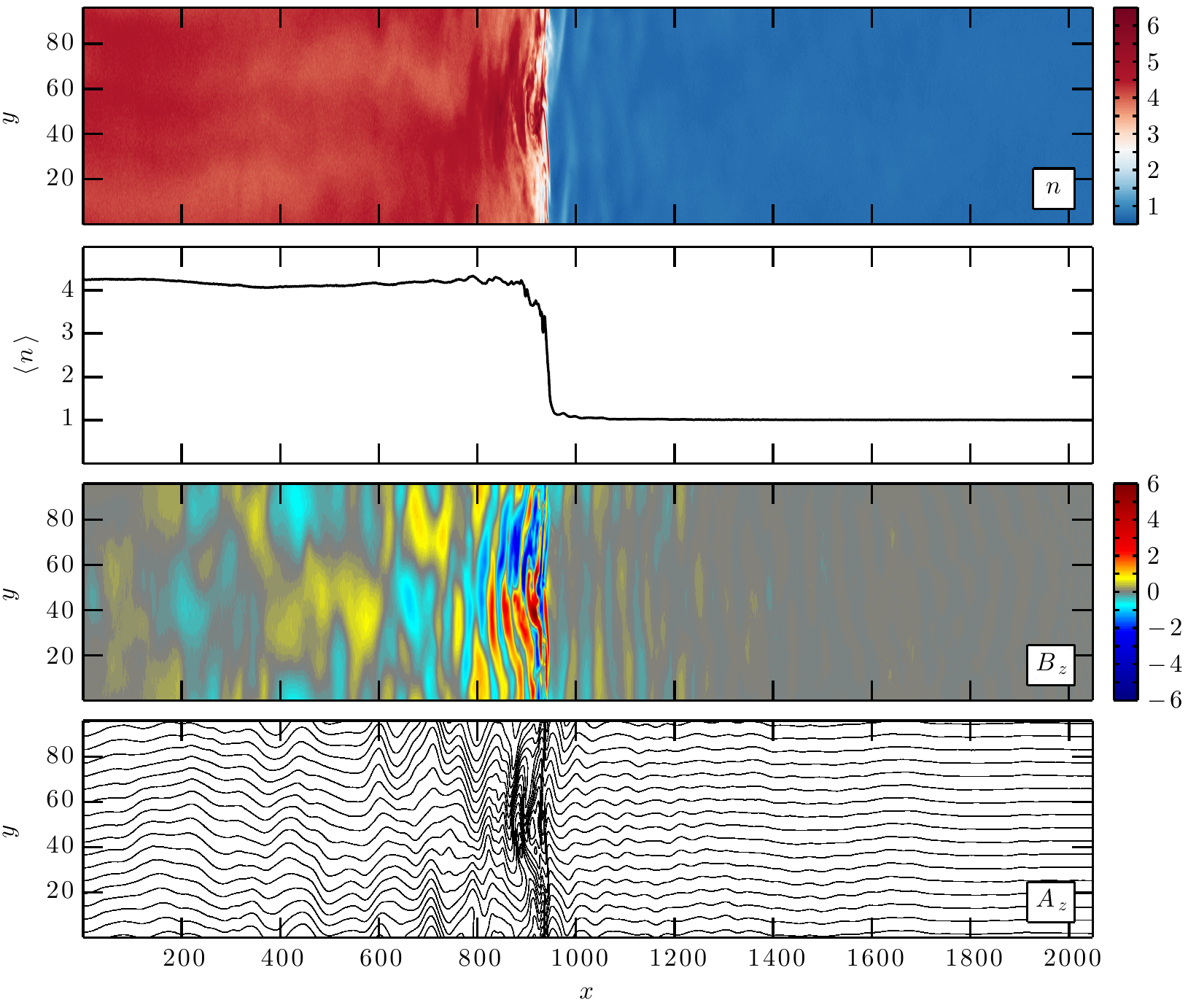}
\caption{Non-relativistic shock test problem using the full-$f$ method. ({\it top to bottom}) Pseudo-color plot of the number density $n$, $y$-averaged number density $\langle n \rangle$ versus $x$, pseudo-color plot of the out-of-plane the magnetic field $B_z$, and contour plot of the magnetic-field lines ($A_z = {\rm const}$) at $t = 300$. The unit of length is the initial ion skin depth, the Alfv\'{e}nic Mach number $v_{\rm sh} = 10$, and $\beta_0 = 2$. See Section \ref{sec:shock} for details.}
\label{fig:shock}
\end{figure}

Non-relativistic shocks constitute a frequent application of fully kinetic and hybrid-kinetic PIC codes \cite[e.g.][]{winske85,quest88,giacalone04,htls07,gs12,cs13,gg13}. Here we demonstrate the ability of our code to simulate such highly nonlinear problems by initializing the distribution function
\[
f(v_x,v_y,v_z) = \frac{n_0}{( \pi \beta_0 )^{3/2}} \exp\left( - \frac{ | \bb{v} + v_{\rm sh} \ex |^2 }{\beta_0} \right)
\]
on a grid of size $L_x \times L_y = 2048 \times 96$ with $8192 \times 384$ cells using 64 particles per cell. The upstream fluid velocity in the downstream reference frame $v_{\rm sh} = 10$, and the initial magnetic field $B_0 = \ex$ is set parallel to the flow with $\beta_0 = 2$. The full-$f$ method is used, and we have set $\eta = 0.5$. Reflecting (outflow) boundary conditions are imposed at $x=0$ ($x = L_x$), while additional particles are injected at $x=L_x$ with speed $v_{\rm sh} = 10$ so that the rightmost cells each contain 64 particles on average. The two counter-propagating flows that result interact to produce a sharp discontinuity, which propagates to the right. This places the downstream fluid at rest, and the kinetic energy of the upstream flow is converted into thermal energy at the shock front. In order to capture the correct entropy jump conditions, the electrons are taken to be adiabatic with $\Gamma_{\rm e} = 2.5$ and $\tau = 1$ so that, after being adiabatically heated at the shock, roughly half of the pre-shock kinetic energy goes into the thermal energy of the electrons. Having to adopt an electron equation of state {\it a priori} is one limitation of studying shock problems with the hybrid-kinetic equations. 

Figure \ref{fig:shock} presents plots of the number density and magnetic field at $t = 300$. From top to bottom, these are a pseudo-color plot of the number density, a plot of the $y$-averaged number density $\langle n \rangle$ versus $x$, a pseudo-color plot out-of-plane component of the magnetic field $B_z$, and a contour plot of the magnetic-field lines ($A_z = {\rm const}$). The shock propagates to the right at speed $\sim$$3$ and is characterized by a density jump of $\sim$$4$, in agreement with the Rankine-Hugoniot jump conditions for a strong shock. Density and magnetic waves are excited at the shock interface and propagate upstream. These waves are compressed at the shock front, leading to a strong self-generated out-of-plane component of the magnetic field that scatters particles and reinforces the shock.

This particular test reveals a well-known issue with numerical integration algorithms that do not employ a total-variation-diminishing, total-variation-bounded, or essentially non-oscillatory scheme for finite differencing: spurious oscillations develop near sharp discontinuities. This is an issue confronting every hybrid-kinetic numerical algorithm that advances the magnetic field by using an electric field computed without suitable upwinding. To remedy this, we have tried upwinding the magnetic field using second-order--accurate van Leer interpolation along the characteristics $\bb{r} = \bb{u} t$ for the calculation of the edge-centered electric field.\footnote{A similar approach has recently been employed by Matsumoto et al.~\cite{mkufs12}.} While this approach preserves monotonicity when advecting sharp features in the magnetic field, it interferes with our code's ability to accurately propagate all wave modes in multiple dimensions. The difficulty with our and other such hybrid-kinetic codes is not so much obtaining an accurate estimate for the time-advanced electric field, but rather calculating the relevant wave characteristics along which to upwind the magnetic field (i.e.~solving the Riemann problem in hybrid-kinetics). In hybrid-kinetic PIC codes, the resulting ringing is customarily tempered by applying a digital filter to smooth the magnetic field \cite[e.g.][]{gbfs07,msmsgp11}. However, in its simplest guise this procedure does not preserve the solenoidal character of the magnetic field. We have instead introduced a diffusive contribution ($\eta \ne 0$) to the electric field, which is used with constrained transport to update the magnetic field in a divergence-free way. Note that the shock problem is the only test problem included in this paper that benefits from a non-zero magnetic diffusivity. We believe that the construction of a monotonicity-preserving integrator for the hybrid-kinetic equations would constitute an important contribution to the field.

\subsection{Axisymmetric kinetic magnetorotational instability}\label{sec:mri2d}

For our final test, we initialize particles in an axisymmetric shearing sheet with the equilibrium distribution function
\[
f(v_x, v_y, v_z) = f_{\rm M}(v) \left( \frac{ 2\varpi + b_z }{ 2 \varpi + b_z - \sigma } \right)^{1/2} \exp\left(-\frac{v^2_y}{\beta_0} \frac{ \sigma }{ 2 \varpi + b_z - \sigma } \right) ,
\]
where $b_z = \cos\theta$ is the component of $\eb$ along the rotation axis. With $\varpi = 0.01$, $\sigma = 0.015$, $\beta_0 = 5$, and isothermal electrons with $\tau = 1$, this setup is unstable to the kinetic magnetorotational instability (KMRI) with growth rate $\sim$$( \sigma \varpi )^{1/2}$ \cite{qdh02}. We adopt these parameters and initiate 1024 particles per cell on a grid of size $L_x \times L_z = \ell/4 \times \ell$ with $N_x \times N_z = 128 \times 512$ cells ($N_{\rm p} \sim 7 \times 10^7$). Here, $\ell$ is the wavelength of the fastest-growing mode as determined by a linear stability analysis (as in ref.~\cite{qdh02}). The resistivity $\eta = 0$, and the full-$f$ method is used.

%
% Figure 10
%
\begin{figure}
\centering
\includegraphics[clip]{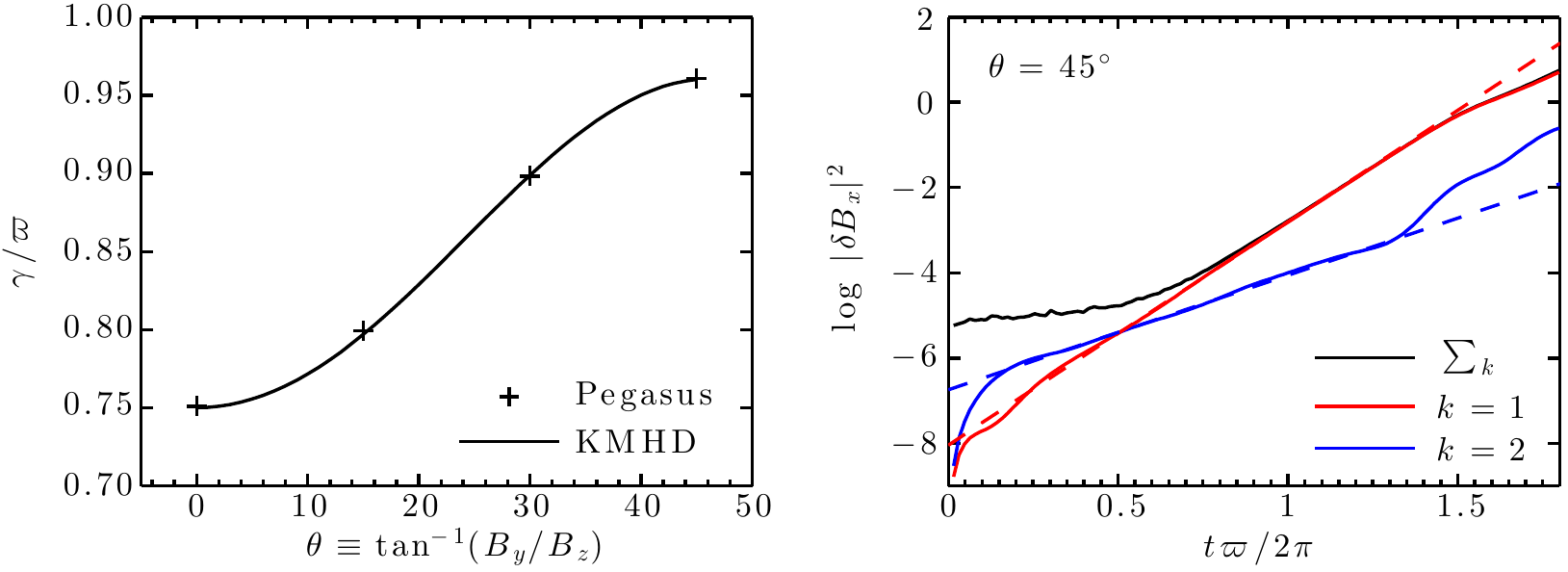}
\caption{Axisymmetric kinetic magnetorotational instability test using the full-$f$ method. ({\it left}) Numerical (plus signs) and analytical (black line) growth rates versus magnetic-field inclination angle, $\theta \equiv \tan^{-1} ( B_y / B_z )$. The analytical growth rates were obtained using kinetic MHD (KMHD) with a collisionless closure \cite{qdh02}. ({\it right}) Temporal evolution of energy in the $k=1$ mode (red solid line), the $k=2$ mode (blue solid line), and all of the modes (black solid line) for $\theta = 45^\circ$. The analytically predicted evolution is traced by the red (for $k=1$) and blue (for $k=2$) dashed lines. See Section \ref{sec:mri2d} for details.}
\label{fig:mri2d-growth}
\end{figure}

We vary $\theta$ between $0^\circ$ (i.e.~vertical magnetic field) and $45^\circ$ (i.e.~inclined magnetic field with $B_y = B_z$) and measure the linear growth rate $\gamma$ as a function of $\theta$. The result is shown in the left panel of Figure \ref{fig:mri2d-growth}; the numerical results (plus signs) match the theoretical results from kinetic MHD (KMHD; solid line) within a fraction of a percent. In the right panel, we show the magnetic-field evolution of select modes for $\theta = 45^\circ$. The numerical results, obtained by Fourier decomposing the simulation domain, are shown for $k = 1$ (red solid line), $k=2$ (blue solid line), and the sum of all the modes (black solid line). The predicted evolution for $k=1$ and $k=2$, obtained from the linear analysis, is traced by the dashed lines. The agreement is excellent up to $\delta B \sim 1$, at which point mirror-mode parasites suppress the positive pressure anisotropy generated by the KMRI by curbing the modes' growth. 

\begin{figure}
\centering
\includegraphics[clip]{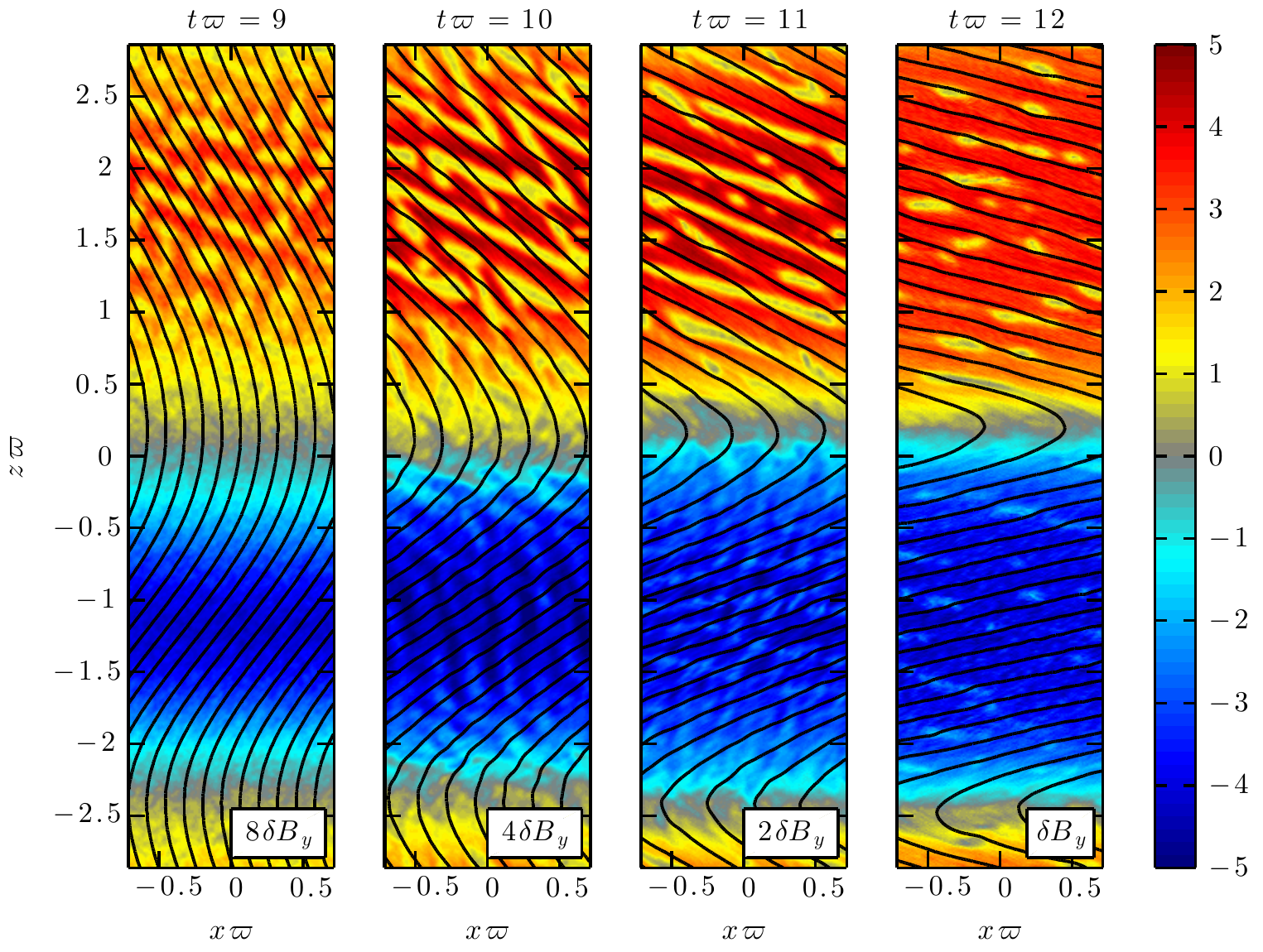}
\caption{Axisymmetric kinetic magnetorotational instability test using the $\delta f$ method. The spatial and temporal evolution of the perturbed azimuthal magnetic field $\delta B_y$ (color) is shown at times $t \varpi = 9,~10,~11,~12$ along with the magnetic-field lines (black lines). The KMRI is characterized by $\delta B_x \delta B_y < 0$, which transports angular momentum outwards. When $\delta B_y \sim 1$, mirror-mode parasites attack the KMRI mode and regulate the pressure anisotropy (see also Fig.~\ref{fig:mri2d-eigen}). Note that the field is rescaled on each plot to exhaust the color-bar limits. See Section \ref{sec:mri2d} for details.}
\label{fig:mri2d}
\end{figure}

\begin{figure}
\centering
\includegraphics[clip]{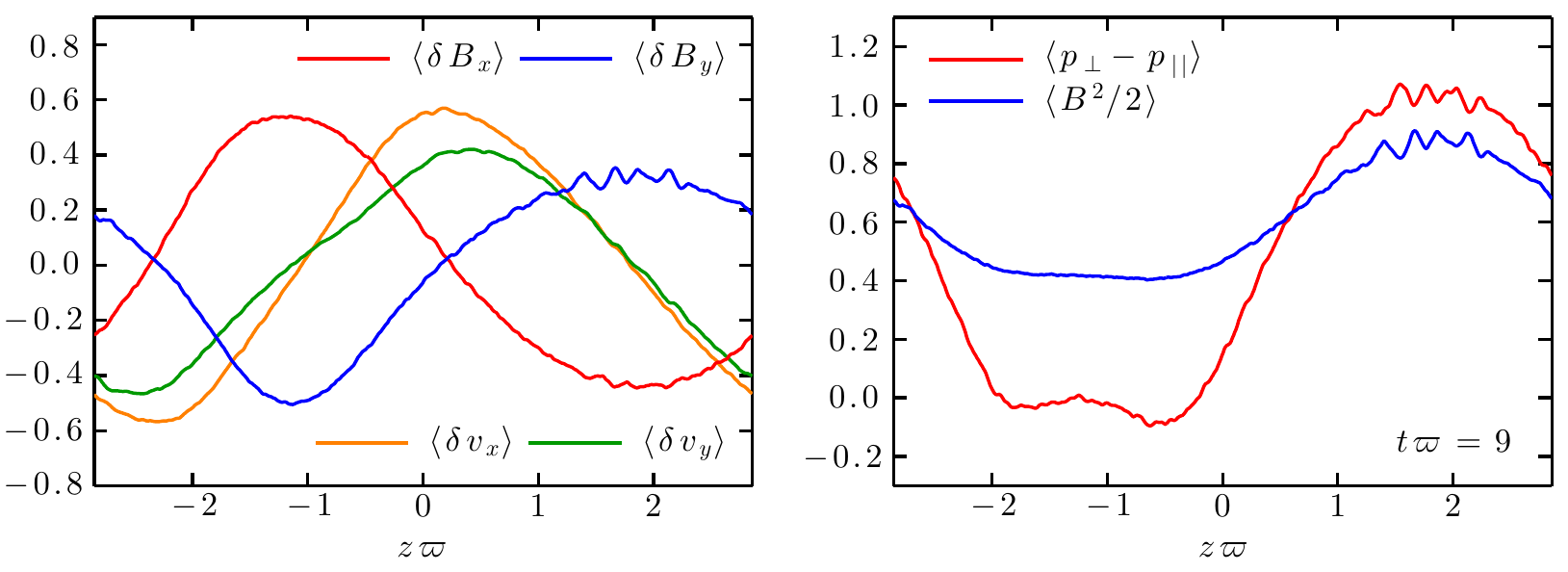}
\newline\newline
\includegraphics[clip]{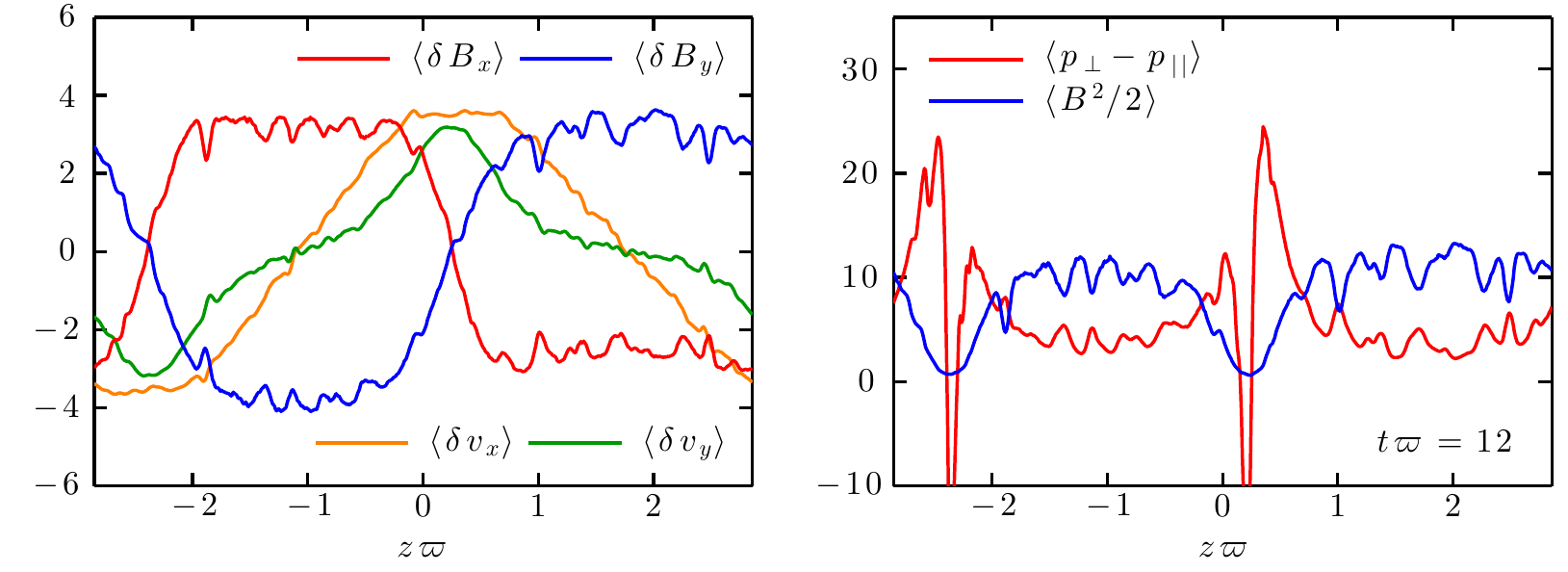}
\caption{Axisymmetric kinetic magnetorotational instability test using the $\delta f$ method. Data is plotted at ({\it top}) $t \varpi = 9$ and ({\it bottom}) $t \varpi = 12$, corresponding to the leftmost and rightmost panels in Figure \ref{fig:mri2d}. ({\it left}) Radially averaged magnetic-field and velocity perturbations. The $x$- (red line) and $y$- (blue line) components of the magnetic field are anti-correlated, while the $x$- (orange line) and $y$- (green line) components of the velocity are correlated. This leads to outward angular-momentum transport. ({\it right}) Regions with $\delta B_{||} = b_y \delta B_y > 0$ generate a positive pressure anisotropy (red line) that exceeds the magnetic energy (blue line) and is unstable to mirror modes with parallel scale $\sim$$20\rho_{\rm i}$. By $t \varpi = 12$, the mirror modes have regulated the pressure anisotropy (note that the pressure-anisotropy and magnetic-energy fluctuations are anti-correlated), and the magnetic (velocity) field has acquired a top-hat- (sawtooth-)like profile. These profiles minimize the effects of pressure anisotropy on the KMRI mode. See Section \ref{sec:mri2d} for details.}
\label{fig:mri2d-eigen}
\end{figure}

This transition to nonlinear behavior can be clearly seen in Figure \ref{fig:mri2d}, which exhibits the perturbed azimuthal magnetic field $\delta B_y$ (color) and the magnetic-field lines (black lines) at times $t \varpi = 9,~10,~11$, and $12$ (i.e.~up to $\sim$$2$ orbits). At $t \varpi = 9$, the magnetic-field lines are approximately sinusoidal, with $\delta B_x \delta B_y < 0$ indicative of outward angular-momentum transport. The amplitude is approaching $\delta B \sim 1$ and mirror-mode parasites begin to emerge. This is presented quantitatively in in the top panels of Figure \ref{fig:mri2d-eigen}, which presents ({\em left}) the horizontally averaged magnetic and velocity fluctuations and ({\em right}) the horizontally averaged pressure anisotropy (compared with the mirror stability threshold $\sim$$\langle B^2/2\rangle$). At this stage, regions with $\delta B_{||} = b_y \delta B_y > 0$ generate a positive pressure anisotropy that  exceeds the magnetic energy and drives mirror-mode parasites. These manifest as small-scale fluctuations with parallel scale $\sim$$20\rho_{\rm i}$. As time progresses, the mirror structures become more pronounced (see the middle panels in Figure \ref{fig:mri2d}), with compressive fluctuations in the magnetic-field lines clearly visible. By $t \varpi = 12$, the magnetic-field lines have become relatively straight except near the midplane, where they change direction at a sharp cusp. The horizontally averaged magnetic and velocity fluctuations and pressure anisotropy at this stage are shown in the bottom panels of Figure \ref{fig:mri2d-eigen}. The magnetic field has acquired a top-hat-like profile, while the velocity exhibits a sawtooth-like profile. These profiles minimize the effects of pressure anisotropy on the KMRI mode (A.~Schekochihin and S.~Cowley, private communication). Over much of the MRI mode, the pressure anisotropy has been reduced below the mirror stability threshold.

The axisymmetric KMRI has also been studied by Riquelme et al.~\cite{rqss12} using a collisionless fully kinetic electromagnetic PIC code. A comparison with their results, as well as a further analysis of the hybrid-KMRI and its long-term evolution in three-dimensional geometry, will be presented in a forthcoming publication [Kunz et al., in preparation].

%
% Figure 11
%

%
% Figure 12
%

\section{Summary}\label{sec:summary}

In this paper we have described {\it Pegasus}, a new hybrid-kinetic PIC code tailored for the study of astrophysical plasma dynamics. {\it Pegasus} incorporates an energy-conserving particle integrator into a stable second-order--accurate three-stage predictor-predictor-corrector algorithm, while relying on the constrained transport method to maintain a divergence-free magnetic field. The code conserves the total magnetic flux through the grid exactly, and offers techniques (e.g.~$\delta f$ method, digital filters) for reducing the impact of finite-number particle noise on the accuracy of the integration algorithm. The effects of rotation and shear are treated locally within the shearing-sheet formalism, with the advection of particles and magnetic fields captured by ``orbital advection"---a second-order--accurate remapping procedure, which removes the orbital motion from the Courant condition, renders the truncation error more uniform in radius, and absolves a finite number of particles from the task of representing what is otherwise an analytically separable piece of the distribution function. {\it Pegasus} is efficiently parallelized to run on multiple processors using domain decomposition through MPI calls, achieving excellent weak scaling out to 4096 cores on a modern high-performance computing cluster.

We have presented a selection of test problems in one, two, and three spatial dimensions that highlight the various components as well as the versatility of the code. These tests demonstrate that {\it Pegasus} is second-order--accurate, captures wave-particle interactions, stably propagates Alfv\'{e}n and whistler waves, reliably handles small-amplitude changes in the distribution function, performs particle and magnetic-field integrations in the shearing sheet, and is able to simulate both linear and nonlinear problems. Further test problems, not presented here, are at forefront of research in space and astrophysical plasma dynamics, and will therefore be the subject of dedicated future publications. These include studies of colllisionless magnetorotational turbulence, perpendicular heating in Alfv\'{e}nic solar-wind turbulence, the saturation of firehose and mirror instabilities in a driven plasma, and the efficiency of particle acceleration in non-relativistic shocks. Some of these problems will necessitate further developments in the code, including a means of driving turbulence and a treatment of ion--ion collisions. The incorporation of these effects and others are eased by the modular design of {\it Pegasus}. 

By simplifying the equations and eliminating nonessential physics from the problem, {\it Pegasus} renders a world of kinetic-scale astrophysical phenomena amenable to rigorous numerical treatment. Our hope is that {\it Pegasus} will become an essential numerical tool of the astrophysical plasma community and, in doing so, will serve to strengthen a growing appreciation for the impact of kinetic-scale plasma physics on the large-scale behavior of the Universe.

\section*{Acknowledgements}

Support for M.W.K. was provided by NASA through Einstein Postdoctoral Fellowship Award Number PF1-120084, issued by the Chandra X-ray Observatory Center, which is operated by the Smithsonian Astrophysical Observatory for and on behalf of NASA under contract NAS8-03060. Support for X.-N.B. was provided by NASA through Hubble Postdoctoral Fellowship Award Number HST-HF-51301.01-A, issued by the Space Telescope Science Institute, which is operated by the Association of Universities for Research in Astronomy for and on behalf of NASA under contract NAS5-26555. The Texas Advanced Computer Center at The University of Texas at Austin provided HPC resources under grant number TG-AST090105, as did the PICSciE-OIT TIGRESS High Performance Computing Center and Visualization Laboratory at Princeton University. This work used the Extreme Science and Engineering Discovery Environment (XSEDE), which is supported by NSF grant OCI-1053575. Aspects of this work were facilitated by the Max-Planck/Princeton Center for Plasma Physics. The authors would like to thank Greg Hammett and Elena Belova for their encouragement, for their suggestions on how to improve our numerical algorithm, and for introducing us to the $\delta f$ method; Anatoly Spitkovsky for fruitful discussions regarding the optimization of PIC algorithms; Damiano Caprioli for his assistance setting up the shock test problem; Lehman Garrison for providing his optimized digital filter algorithm; Joshua Dolence and Ammar Hakim for their advice with coding issues; Tobias Heinemann for useful conversations regarding the shearing-sheet formalism in the context of the hybrid-kinetic equations; and Alexander Schekochihin and Steven Cowley for sharing their expertise on multiscale plasma kinetics and inspiring much of the work presented here.

\bibliographystyle{unsrt}

\bibliography{ksb13}

\end{document}